\documentclass[]{aa}
\usepackage[varg]{txfonts}
\usepackage{graphicx}
\usepackage{booktabs}
\usepackage{amsmath}
\usepackage{colortbl}
\usepackage{subfig}
\usepackage{rotating}
\usepackage{amssymb}
\usepackage{latexsym}
\usepackage{ifthen}
\usepackage{enumitem}

\begin{document}

\title{SINFONI spectra of  heavily obscured AGNs in COSMOS: evidence of outflows in a MIR/O target at z$\sim2.5$\thanks{Based on observations with SINFONI VLT spectrograph, ESO program 092.A-0884(A)}}

\author{M. Perna
		\inst{\ref{i1},\ref{i2}}\thanks{E-mail: michele.perna4@unibo.it}
		\and 
	M. Brusa
		\inst{\ref{i1},\ref{i2}} 
		\and
	M. Salvato
		\inst{\ref{i3},\ref{i4}}
		\and
	G. Cresci
		\inst{\ref{i5}} 
	\and
	G. Lanzuisi
		\inst{\ref{i1},\ref{i2}}	
		\and
	S. Berta
	        \inst{\ref{i3}}	
		\and
	I. Delvecchio
                \inst{\ref{i6}}
                \and
        F. Fiore
                \inst{\ref{i7}}
                \and
        D. Lutz
                \inst{\ref{i3}}
	\and
	E. Le Floc'h
		\inst{\ref{i8}}   
  	        \and
	V. Mainieri
		\inst{\ref{i9}} 
	\and
        L. Riguccini
	        \inst{\ref{i10}}
}

%\offprints{M. Perna, \email{michele.perna4@unibo.it}}

\institute{Dipartimento di Fisica e Astronomia, Universit\`a di Bologna, viale Berti Pichat 6/2, 40127 Bologna, Italy\label{i1}
	\and
	INAF - Osservatorio Astronomico di Bologna, via Ranzani 1, 40127 Bologna, Italy\label{i2}
	\and
	Max Planck Institut fur Extraterrestrische Physik, Giessenbachstrasse 1, 85748 Garching bei M\"unchen, Germany\label{i3}
	\and
        Excellence Cluster Universe, Boltzmannstrasse 2, D-85748 Garching bei M\"unchen, Germany\label{i4}
	\and
	INAF - Osservatorio Astrofisico di Arcetri, Largo Enrico Fermi 5, 50125 Firenze, Italy\label{i5}
	\and
       Department of Physics, University of Zagreb, Bijeni\v{c}ka cesta 32, HR-10000 Zagreb, Croatia\label{i6}
	\and
	INAF - Osservatorio Astronomico di Roma, via Frascati 33,   00044 Monte Porzio Catone (RM), Italy\label{i7}
	\and
	AIM, Unit\'e Mixte de Recherche CEA CNRS, Universit\'e Paris VII, UMR n158, 75014 Paris, France\label{i8}
	\and
	European Southern Observatory, Karl-Schwarzschild-str. 2,  85748 Garching bei M\"unchen, Germany\label{i9}
	\and
	Observat\'orio do Valongo, Universidade Federal do Rio de Janeiro, Ladeira do Pedro Ant\^onio 43, Sa\'ude, Rio de Janeiro, RJ 20080-090, Brazil\label{i10}
}

\date{Received 2 November 1992 / Accepted 7 January 1993}

\abstract {} 
{We present new data for four candidate obscured Compton-Thick (CT) quasars at z $\sim$1-2.5 observed with SINFONI VLT spectrograph in AO mode. These sources were selected from a 24$\mu$m Spitzer MIPS survey of the COSMOS field, on the basis of red mid-infrared-to-optical and optical-to-near-infrared colours, with the intention of identifying active galactic nuclei (AGNs) in dust enshrouded environments, where most of the black hole mass is assembled in dust enshrouded environments.} 
{ Near infrared spectra were analyzed in order to check for emission
  line features and to search for broad components in the
  [OIII]-H$\beta$ and H$\alpha$-[NII] regions.  X-ray spectral analysis, radio and MIR diagnostics, and SED fitting have also been employed to study the nature of the sources.}
{We successfully identified three objects for which we had only a photometric redshift estimate. Based on their emission line diagnostics and on ancillary multi-wavelength constraints, we find that all four targets harbor obscured AGNs. Broad profiles that could be attributed to the effects of outflows are revealed in only one target, MIRO20581.
In particular, we clearly resolved a fast ($\sim$1600 km/s) and extended ($\sim$5 kpc) outflow in the [OIII]5007 emission line. 
This feature, the commonly used indicator for ionised outflowing gas,
was sampled and detected only for this target; hence, we can not exclude the presence of outflows in the other sources.
Overall, the constraints we obtain from our targets and from other comparative samples from the literature suggest that these optically faint luminous infrared galaxies, hosting obscured AGNs, may represent a brief evolutionary phase between the post-merger starburst and the unobscured QSO phases.
} 
{} 

\maketitle
\titlerunning{Evidence of outflows in a MIR/O target at z$\sim2.5$} 

\section[Introduction]{Introduction}

Outflow winds are predicted to be ubiquitous in active galactic nuclei (AGN) systems and are invoked in many co-evolutionary models to link the growth of  supermassive black hole (hereafter BH) and galaxies through feedback phenomena. 
These models predict an obscured phase for young recently ignited quasars, triggered by the funneling of a large amount of gas into the nuclear region during major galaxy mergers \citep[e.g.,][]{Menci2008,Hopkins2008}. Roughly at the same time, this amount of gas is also responsible of vigorous star-formation activity. This initial phase is followed by a transitional phase, the so-called feedback or blow-out phase (\citealt{Hopkins2008}), in which the gas is cleared out through outflowing winds released by the BH, before becoming a normal unobscured quasar (QSO).  

In the framework described above, during the obscured phase, the BH is expected to accrete mass very rapidly, implying a vigorous, although obscured, X--ray emission. In the census of AGNs, X--ray surveys have been extensively used to probe the assembly and growth of BH at high redshift. In particular, since the X--ray flux is less attenuated than the optical flux, selection criteria based on high X--ray-to-optical flux ratio ($f_x/f_O$) have been used to select obscured sources at z $\sim1-2$. Several studies \citep[e.g.,][]{Fiore2003, Mignoli2004, Alexander2002, DelMoro2009, DellaCeca2015} have found that sources with high $f_x/f_O$ are characterized, on average, by red optical-to-near-infrared colours ($R-K_{Vega}>5$) and column densities in the X--ray of the order of 10$^{21-23}$cm$^{-2}$.  
Moreover, VLT XSHOOTER (\citealt{B15,P15}) and SINFONI (\citealt{Cresci2015}) observations of a small sub-sample of obscured QSOs at z$\sim$1.5, selected on the basis of their observed red colours and high $f_x/f_O$ ratio, have confirmed the presence of ionised outflowing material in 75\% of object and a dust-reddened type 1 nature.
\begin{figure}[!t]
\centering
\includegraphics[width=9cm,angle=0]{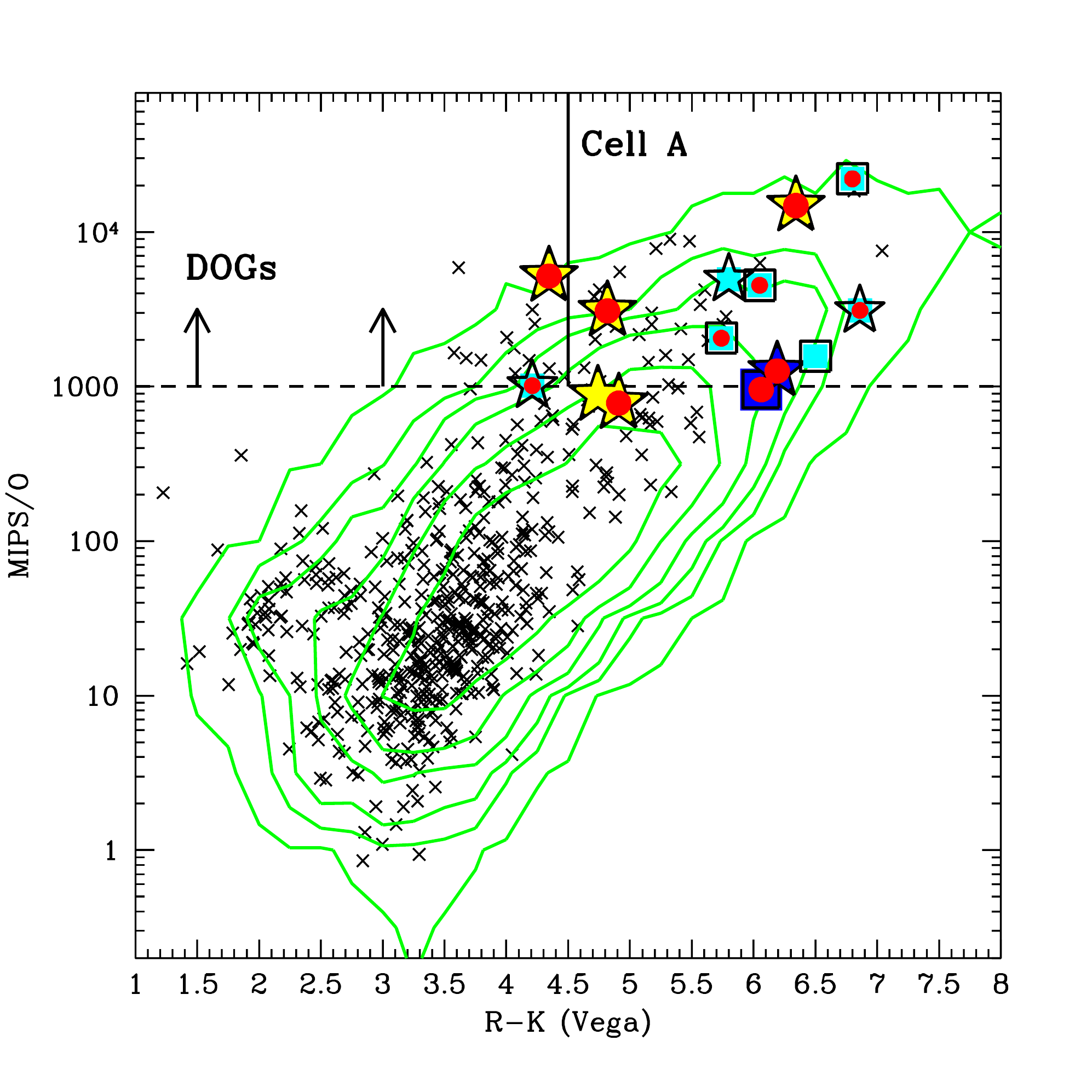}
\caption{$f_{24 \mu m}/f_R$ as a function of $R-K$ colour for all the
  COSMOS 24 $\mu$m sources associated with the optical and K-band
  counterparts (green isodensity contours). Black crosses represent
  the 1 mJy flux-limited sample. Yellow stars denote the 5 objects
  observed with SINFONI; cyan stars and squares denote
  \citet{Melbourne2011} and \citet{Brand2007} targets, respectively
  (two targets present in both samples are indicated with stars); blue
  symbols denote the \citet{P15} targets. Star and square symbols
  indicate IFU and long-slit observations, respectively. Sources
  marked with red circles are also detected in the 1 mJy flux limited
  sample and are those considered for the analysis. The box delimited
  by vertical solid line and horizontal dashed line marks the  region
  proposed in \citet{Fiore2009} to select CT AGN (Cell A). The
  horizontal dashed line sets the criterion for the selection of DOGS.}
\label{mirofiore}
\end{figure}
These objects appear similar to infrared bright (K$_{Vega}\le$16) dust-reddened Type 1 QSOs selected by combining radio with near-infrared (NIR) and optical catalogues at lower redshift (z $\lesssim 1$; \citealt{Glikman2004,Glikman2007}). Studying a sub-sample of 13 objects, with strongly disturbed morphology, 
\citet{Urrutia2012} found that $\sim$60\% of these radio-detected sources show evidence for outflow in the [OIII]5007 line profile (see also \citealt{B15}).     

However, each selection technique biases the samples towards particular properties and, chiefly, may biases the characterization of the outflows (see \citealt{B15} for details); 
to test co-evolutionary models, we need to select and isolate
different populations of quasars in the different phases of the
AGN-galaxy co-evolution, including the initial Compton-Thick (CT\footnote{ Throughout the paper, we distinguish between moderately
obscured (with log(N$_{\rm H}$)=22-23 cm$^{-2}$), highly obscured (with log(N$_{\rm H}$)=23-24 cm$^{-2}$), and CT AGNs
(with log(N$_{\rm H})>$24 cm$^{-2}$).})
phase.

Since most of the absorbed AGN energy is re-emitted in the mid-infrared (MIR), surveys at these wavelengths can potentially recover the elusive obscured accretion missed by X--ray surveys \citep[e.g.,][]{Brandt2015}.
Several criteria based on the MIR emission of high-z sources have been
introduced in the recent years to search for heavily obscured AGNs at
z $\sim1-3$, and  have been applied on  Spitzer MIPS observations in
multiwavelength survey fields. Typically, the criteria involve the
selection  of objects with MIR luminosities typical of AGN but with
faint optical or near-infrared emission \citep[e.g.,][]{Martinez2005,
  Fiore2008, Fiore2009, Dey2008, Riguccini2015}.  

For example, \citet{Fiore2009} used the MIPS 24$\mu m$ COSMOS catalogue (\citealt{Sanders2007}) to select a sample of $\sim60$ candidate obscured AGN/CT QSOs characterized by extreme mid-infrared-to-optical flux ratio ($f_{24\mu m}/f_R>1000$) in the area covered by the C-COSMOS Chandra survey (\citealt{Elvis2009,Civano2012}). They coupled this selection with a red colour $(R-K)_{Vega}>4.5$ cut, which corresponds efficiently pick up objects at  the redshift of interest (z $\sim1-3$). 
To test the efficiency of the selection, they stacked the Chandra images at the position of the MIPS sources without a direct X--ray detection and recovered a  hardness ratio (HR) in the stacking signal larger than that measured for less extreme sources ($f_{24\mu m}/f_R<1000$ and/or $(R-K)<4.5$). 
Still, these sources with $f_{24\mu m}/f_R>1000$ exhibit evidence of both star formation and AGN activity, and there are contradicting conclusions on how many of the sources selected in this way are actually obscured AGN at z $\sim1-3$, rather than dusty star forming objects \citep[e.g.,][]{Donley2008, Fiore2009, Dey2008}.

In this paper we present new SINFONI observations, assisted with
Adaptive Optics (AO), for a sample of four luminous, highly obscured QSOs in the COSMOS field, selected on the basis of their high mid-infrared-to-optical flux ratios (MIR/O) and red R-K colours. 
 Given the tight positive correlation between the $f_{24\mu m}/f_R$ and the
$f_x/f_O$ ratios found  for sources with column density of the order
of 10$^{22-23}$ cm$^{-2}$ (see \citealt{Fiore2008}, Fig.2), \citet{Fiore2008}
suggested that luminous highly obscured AGNs, that are
faint in the X--ray because of the high column densities,
i.e. N$_H\gtrsim$ 10$^{23-24}$ cm$^{-2}$,  and that cannot be selected
using their X/O ratio, can be recovered using their MIR/O ratio. Therefore, potentially, in the framework previously described, we may be able to select sources in the prelude, or at the beginning, of the blow-out phase.

In the following we will refer to these sources as MIRO targets, as per their high MIR/O flux ratio, and using their MIPS catalogue ID from the COSMOS Spitzer catalogue (\citealt{LeFloch2009}). 
The main aim of the SINFONI observations, besides the spectroscopic
determination of the redshifts of the targets, is to compare the
physical properties of the sources selected in different ways (MIR/O
vs. X/O) and assess the presence of ionised outflows and broad
features. Targets selected with a simple MIR/O ratio cut are usually
known in the literature as dust obscured galaxies (DOGs,
\citealt{Dey2008}; see also \citealt{Riguccini2011,Riguccini2015} for
a complete discussion on the DOGs population in COSMOS). We will
compare our results with a compilation of few DOGs sharing similar
properties, e.g., redshift, 24$\mu m$ flux, in the last Section.  

The paper is organised as follows: Sect. 2 presents the sample selection and the ancillary data collected for our MIRO targets; Sect. 3 outlines the VLT observations and data reduction; Sect. 4 exposes the spectroscopic analysis. Sect. 5 presents proof of ionised outflowing material in the X--ray source MIRO20581 and discusses the energetic output associated to the outflow and finally we summarise our results and the implications in Sect. 6. Throughout the paper, we adopt the cosmological parameters $H_0$=70 km/s, $\Omega_M$=0.3 and $\Omega_\lambda$=0.7 (\citealt{Spergel2003}).  We adopt a Chabrier initial mass function to derive stellar masses and star formation rates (SFRs).

\begin{table*}
\footnotesize
\begin{minipage}[!h]{1\linewidth}
%\centering
\caption{MIRO/SINFONI sample: selection properties; log file of observations}
\begin{tabular}{lcccccccc|cccc}
 MIRO &RA & DEC &z$_{phot}$ & R   & K   &f$_{24\mu m}$  & R-K  & MIPS/O &band &Guide Star& expo & $z_{spec}$\\
      &   &     &             & AB  & AB  & mJy        & Vega    &
      &&(name)&(min)&(this work)\\
(1)      & (2)& (3)   & (4)              & (5)  & (6) & (7) & (8) &
(9) & (10) & (11) & (12) & (13)\\
\hline
18744 &10:01:52.2 & 01:56:08.6    & 0.97$^a$ & 22.8&19.6 &2.08$\pm$0.03 &   4.90 &  780& J  &Hip037044 &  50 &0.97 \\
10561 &09:59:43.5 & 01:44:07.6  & 1.54 & 24.6&21.4 &1.70$\pm$0.06 &  4.80  & 3090& J,H&Hip040661 & 50,50 &1.43   \\ 
28704 &10:01:45.9 & 02:28:53.8  & 1.74 & 26.3&21.6 &1.63$\pm$0.02 &  6.35  & 14800& HK &Hip046054  & 30 &1.64 \\
20581 &10:00:00.6 & 02:15:31.1  & 2.09 & 25.3&22.6 &1.47$\pm$0.02 &   4.35 & 5180& HK&Hip046054  & 80 &2.45  \\
18433 &10:01:44.8 & 01:55:55.8  & 2.59 & 24.6&21.6 &0.44$\pm$0.02  &   4.75 & 880& HK &Hip044598  & 120& ---  \\
\hline
\end{tabular}
\label{sample}
\end{minipage}
Notes: {(1) target name; (2) right ascension; (3) declination; (4) photometric
redshifts available prior to the SINFONI observations; (5) and (6): R
and K-band magnitudes; (7) MIPS 24$\mu m$ flux; (8) R-K colour; (9)
MIPS 24$\mu m$/O flux ratio; (10) SINFONI filters; (11) guide star
name; (12)  total integration time on-target for each band; (13) spectroscopic redshift.}\\
$^a$ Spectroscopic redshift available from Magellan IMACS spectrum.
\end{table*}

\begin{table*}
\footnotesize
\begin{minipage}[!h]{1\linewidth}
%\centering
\caption{MIRO/SINFONI sample: main properties}
\begin{tabular}{lcccccccccccccc}
 MIRO   & N$_{\rm H}$         & log(L$_{\rm 2-10}$) & log(L$_{5.8}$) & log(L$_{\rm bol}$) & S$_{\rm radio}$ & q$_{24}$&M$_*$& SFR&log(R$_{SB}$)$^b$ &E(B-V)$_{host}$&E(B-V)$_{AGN}$\\
        & 10$^{23}$ cm$^{-2}$ & erg s$^{-1}$   & erg s$^{-1}$ & erg s$^{-1}$  &  $\mu$Jy     &    & 10$^{11}$M$_{\odot}$&M$_{\odot}$/yr& & &   \\
(1) & (2) & (3) & (4) & (5) & (6) & (7) & (8) & (9) & (10) & (11) & (12)\\
\hline
18744   & 2.4$_{-1.3}^{+7.5}$ &  43.82   & 44.85 & 45.11 & 424$\pm$28   & 2.3 &2.0& 99     &0.2&0.5& 2.7 \\
10561   & 2.9$_{-1.3}^{+1.8}$ & 43.96    & 45.35 & 45.97 & 72$\pm$15  &0.4 & 1.5 &196&0.3     &0.7 & 4.0 \\
28704   & ---                 & ---      & 45.32 & 47.63 & 154$\pm$25   & 2.0 &0.6&11($<$25)$^a$&-0.8($<$-0.5)$^a$&0.5& 4.1 \\
20581   &  6.8$_{-2.1}^{+3.0}$& 45.00    & 45.89 & 46.61 & 5430 $\pm$60 &-0.6 &1.9&48($<$132)$^a$&-0.6($<$-0.2)$^a$&0.6&2.0\\
%18433 ({\bf remove?})   & ---                 & ---            & -- & -- &  ---        &---  &---&---&---&---&---\\
\hline
\end{tabular}
\label{properties}
\end{minipage}
Notes: (1) target name; (2) column density; (3) intrinsic X-ray
luminosity; (4) rest-frame 5.8$\mu m$ luminosity; (5) bolometric
luminosity; (6) rest-frame 1.4 GHz flux; (7) q$_{24obs}=log(f_{24\mu
  m}/f_{1.4GHz})$ (\citealt{Bonzini2013}); (8) stellar mass; (9) star
formation rate; (10) starburstiness, defined as the ratio between the
specific Star Formation Rate (sSFR=SFR/M$_*$) and that expected for
Main Sequence galaxies  at given computed stellar mass and
spectroscopic redshift (sSFR$_{MS,z}$), according to the relation of
\citet{Whitaker2012}; (11) galaxy reddening ; (12) AGN reddening.\\
$^a$ Values in the parenthesis refer to measurements constrained using upper limits in the FIR SED (see Figure \ref{SED}). 
\end{table*}

\section{Sample selection}

\begin{figure}
\centering
\includegraphics[width=9cm]{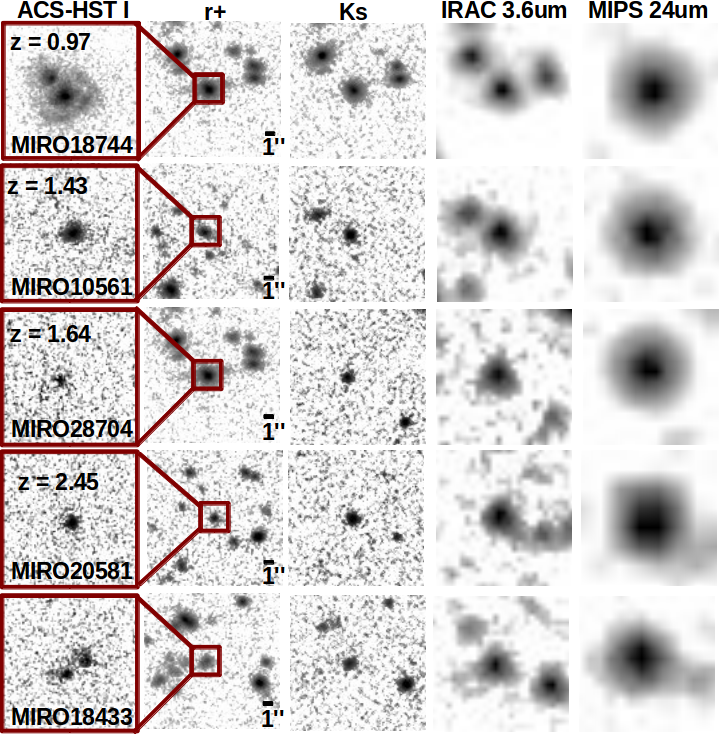}
\caption{From left to right: ACS-HST I, r+ SUBARU, Ks COSMOS, IRAC 3.6 $\mu$m, MIPS 24 $\mu$m band cutouts of the five MIRO targets. 
The target name, the position, the cutouts scale  and the redshift derived from the SINFONI data are also labelled. For display purposes, 3\arcsec x3\arcsec ACS-HST cutouts  show the regions in the red boxes superimposed on the r+ cutouts.}
\label{cutouts}
\end{figure}

Figure \ref{mirofiore} shows the mid-infrared-to-optical flux ratio
$f_{24\mu m}/f_R$ versus the (R-K) colour diagnostics diagram proposed
in \citet{Fiore2008}, applied to the MIPS-selected sources in the
COSMOS field (\citealt{Fiore2009}). The green isodensity contours show
the distribution of the full sample of MIPS-selected sources in the
COSMOS GO3 data (\citealt{LeFloch2009}) and with associated optical
and K-band counterparts ($\sim15000$ sources), while the black crosses
are the sources detected at fluxes larger than 1 mJy at 24$\mu m$
($\sim550$). The box in the top right corner in the colour-colour space, delimited
  by vertical solid line and horizontal dashed line, mark the
region where CT AGN are expected in more than 60\% of the
MIPS-selected sources (cell A in Fiore et al. 2009). All the sources
above the horizontal dashed line are instead usually referred as  DOGs in the literature.

From the \citet{LeFloch2009} MIPS selected sample, we observed with
SINFONI-NGS mode five targets marked as yellow stars in Figure
\ref{mirofiore}, and listed in Table \ref{sample}, with the fluxes in
the R, K and MIPS 24$\mu$m band used for the selection. Three out of
five are classified as DOGs (MIRO10561, MIRO28704, MIRO20581) and 2
out of five are also in Cell A of Fiore et al. (2009; MIRO10561,
MIRO28704)\footnote{ At the time of observations, all the
  targets were selected within the cell A. Differences from the
  current situation shown in Figure \ref{mirofiore} are due to the
  fact that we now use an improved version of the COSMOS photometric catalogue (Laigle et al. 2015 in preparation) in which the photometry is slightly changed. The colours in the figure are all related to the
``total flux'' measurements.}.
The main selection criteria, in addition to the high MIPS/O fluxes and the red R-K colour, were the proximity to a bright AO star and a photometric redshift broadly in the range z$\sim1-3$ so that rest-frame optical lines are redshifted in the SINFONI J, H or K filters (four targets). In addition,  for only one source (MIRO18744) a spectroscopic redshift was available from the IMACS/Magellan follow-up of X--ray sources in the COSMOS field (Trump et al. 2007), and we proposed to observe its H$\alpha+$[NII] region with SINFONI (J band).

In Figure \ref{mirofiore} we  also show the z $\sim2$ DOGs presented
in Melbourne et al. (2011; cyan stars) and in Brand et al. (2007; cyan
squares) for which K-band magnitude is available, and the z $\sim1.5$
dust-reddened type 1 sources presented in Perna et al. (2015; blue symbols). The properties of these targets will be discussed in Section \ref{discussion}, in order to compare our results with those previously reported  in literature. 

The ACS (3\arcsec x 3\arcsec), R-band, K-band, IRAC3.6$\mu$m and MIPS
(15\arcsec x15\arcsec) cutouts\footnote{The cutouts are extracted from
  the public COSMOS cutouts
  page:http://irsa.ipac.caltech.edu/data/COSMOS/index\_cutouts.html}
of the 5 targets are shown in Figure \ref{cutouts}. We included
also the IRAC 3.6$\mu$m cutout in order to verify blending problems in
the MIPS emission: for all but MIRO18433, we can safely say that the majority of
                              the emission at 24 micron is correctly
                              associated to the K-band and optical
                              counterpart (e.g. our SINFONI targets),
                              and that the observed colours do not
                              suffer from bad photometry.
                                MIRO18744 may show evidence for an
                                ongoing merger: tidal tails and double
                                nuclei are distinguishable in the ACS
                                cutout.
MIRO18433, instead, presents two components in the ACS cutout (last row of Figure \ref{cutouts}), which are strongly blended at optical and infrared wavelengths and preclude an accurate SED fitting decomposition and a correct photometric redshift derivation.  Indeed, MIRO18433 is the only source for which no spectral features have been detected in the SINFONI HK band in 2 hours observations (see Table \ref{sample}). Moreover, MIRO18433 is the only source below the 1mJy flux limited sample (red circles in Figure 1).
For all the above reasons, MIRO18433 was excluded from the subsequent analysis.

\begin{figure}
\centering
\includegraphics[width=8cm,angle=0]{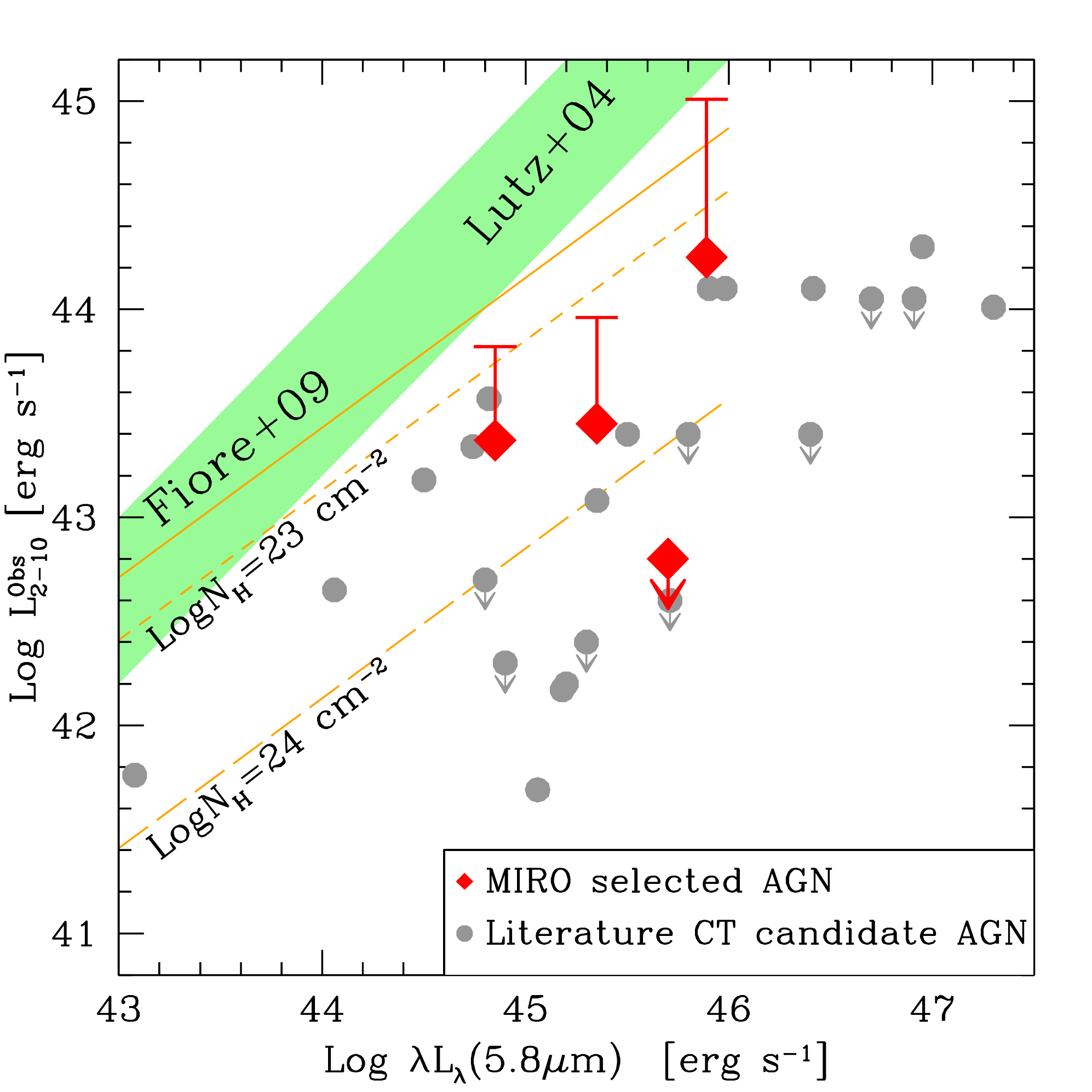}
\caption{Log(L$^{Obs}_X$ ) vs Log(L$_{5.8\mu m}$) for several CT
  candidates. Grey circles represent literature CT candidate AGN (see
  \citealt{Lanzuisi2015a} for more details). Red diamonds represent
  our MIRO targets; intrinsic X--ray luminosities of the X--ray
  detected objects are also indicated with upper bars. The green
  shaded area is the relation of Lutz et al. (2004) for a sample of
  low redshift unobscured AGN. The orange solid line is the relation
  for high redshift unobscured AGN (Fiore et al. 2009), while the dashed and long dashed lines are the expected relation for a 10$^{23}$ cm$^{-2}$ and 10$^{24}$ cm$^{-2}$ absorber.}
\label{l58lx}
\end{figure}

\begin{figure*}
\centering
\includegraphics[width=5.3cm,angle=0]{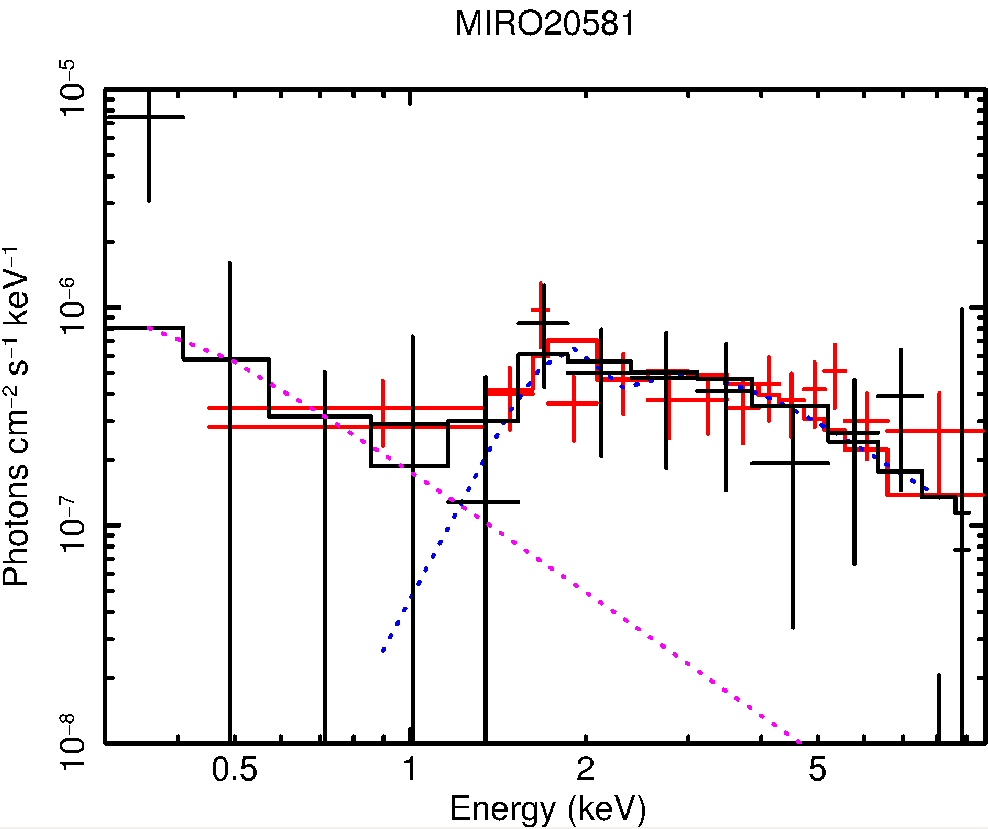}
\includegraphics[width=5.3cm,angle=0]{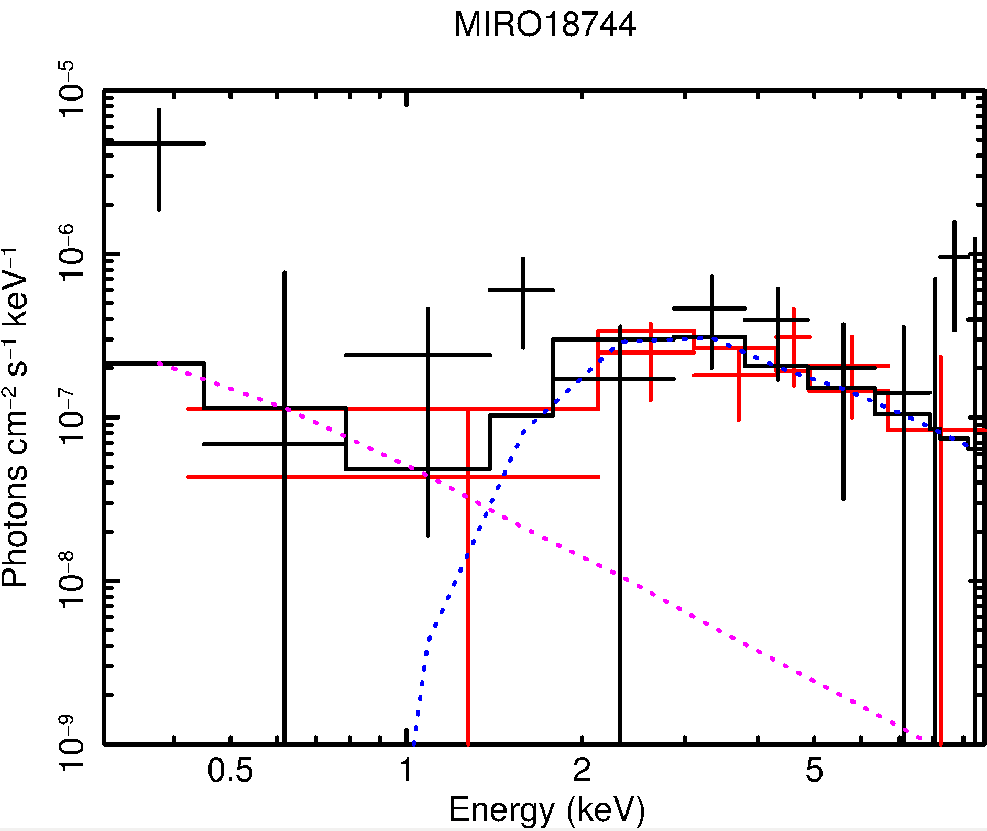}
\includegraphics[height=4.5cm,angle=0]{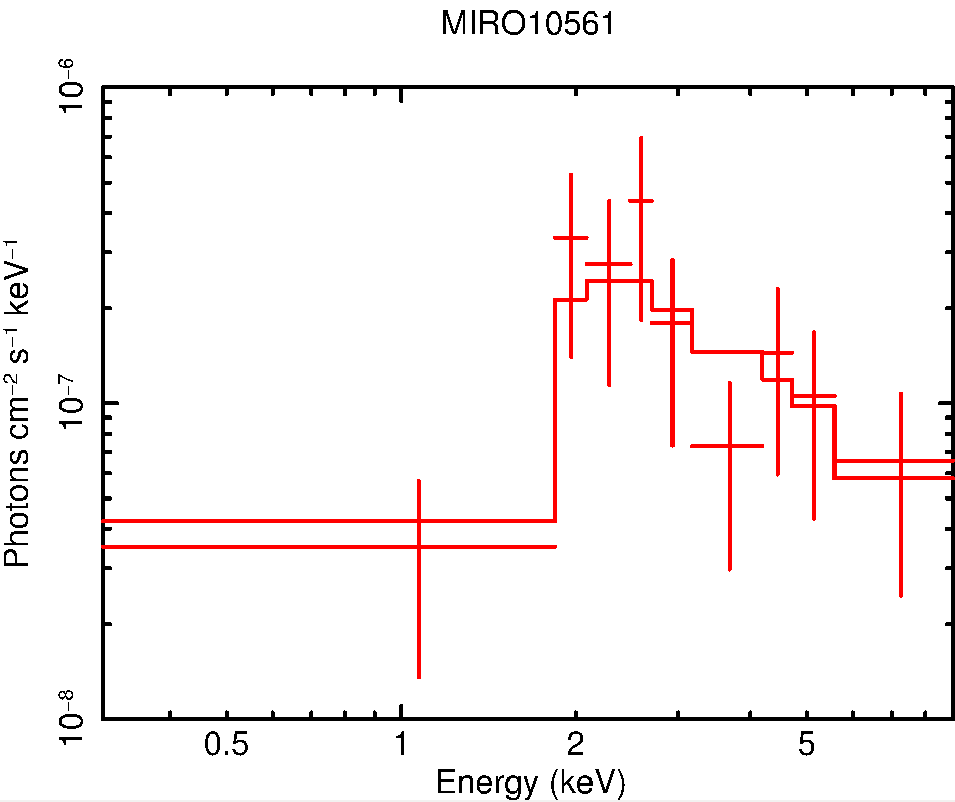}
\caption{X--ray spectra of MIRO20581 (left), MIRO18744 (centre)  from XMM and Chandra, and MIRO10561 (right) from Chandra. The XMM (black) and Chandra (red)  data of MIRO20581 and MIRO18744 are fitted with a double power-law (as shown in the model components). The MIRO10561 {\it Chandra} data are instead fit with a single absorbed power-law.}
\label{MIROspectraXMM}
\end{figure*}

\subsection{Identikit via ancillary data}\label{multiwav}
Previous works (e.g. Alexander et al. 2002, Donley et al. 2008) have demonstrated that red optical-to-near-infrared colours and high MIR-optical ratios can identify both AGN and star-forming galaxies. 
In the following, we briefly discuss the multiwavelength properties of
the SINFONI targets in order to assess which one among SF or AGN
activity is the dominating process. We note however that all our {\bf four} targets have $f_{24\mu m}>$ 1 mJy, and that the 24$\mu$m emission is, on average, increasingly dominated by AGN contribution at higher  $f_{24\mu m}$ \citep[e.g.,][]{Brand2006,Dey2008}. In the subsequent analysis (e.g. X--ray and SED fits) we made use of the spectroscopic redshift obtained from our SINFONI observations.  \\ 
\begin{enumerate}
\item 
%\noindent
 {\it MIR-X--ray diagnostics}: The AGN intrinsic hard X--ray luminosity and the infrared luminosity re-emitted by the torus follow a tight correlation (\citealt{Lutz2004, Gandhi2009}). 
Figure \ref{l58lx} shows the distribution of rest-frame observed
X--ray luminosity (L$^{Obs}_X$) vs. L$_{5.8\mu m}$  for several
samples of CT candidate AGN\footnote{All the sources having upper
    limits to the X-ray luminosity have been pre-selected as AGN
    candidate using a variety of methods specifically designed to
    discriminate between SF and AGN galaxies.} collected by \citealt{Lanzuisi2015a} (with the addition of XMM ID 5371; \citealt{Civano2015}) and for our MIRO targets (red diamonds). \\
For all the sources, the rest-frame 5.8$\mu m$ luminosities were obtained using a simple power-law interpolation  between the 24 and 8$\mu$m observed-frame luminosities. The X--ray luminosities have been computed on the basis of the available XMM and {\it Chandra} data in the COSMOS field. More in details, MIRO20581 and MIRO18744 are both detected in the X--rays, in the XMM-COSMOS (XID70135 and XID60205: \citealt{Cappelluti2007,B10}) and C-COSMOS (CID451 and CID401; Civano et al. 2012) surveys, while MIRO10561 is detected in the deeper COSMOS-Legacy survey (CID 3587 in the Civano et al. 2015 catalogue and Marchesi et al. in preparation). The remaining source, MIRO28704, is instead undetected down to a luminosity of log(L$_{X}$)$\sim42.8$ in the 2-10 keV band. \\
The Lutz et al. (2004; green shaded area) and Fiore et al. (2009;
orange solid line)  relations represent the tight X--ray-to-mid-IR
correlations found for the low- and high-redshift ({\bf z$>$1}) unobscured AGN,
respectively. The Lutz et al. (2004) relation has been confirmed
recently also at higher redshift by \citet{Mateos2015}. These
relations, however, have been calibrated at low luminosities, and require
extrapolations to high luminosities (i.e. log$(\lambda
L_{\lambda}(5.8\mu m))>46$). A flattening of the MIR-Xray relation at the highest luminosities has been found by the recent work of \citet{Stern2015}. 
Assuming that both the hard X--ray and the infrared luminosities are related to the AGN activity (see below; Figure \ref{SED}), given that the mid-IR is largely independent of obscuration, a lower L$_X$ to L$_{5.8}$ ratio with respect to that observed for unobscured AGN suggests that the observed L$_X$ are affected by obscuration (e.g. \citealt{Stern2014}).
All the MIRO sources lie below these relations, at values consistent
with a heavily obscured absorber: the long- (short-)	dashed line in
Figure \ref{l58lx} marks the N$_{\rm H}=10^{24}$ (10$^{23}$) cm$^{-2}$
locus, computed from the Fiore et al. relation. Therefore, the N$_{\rm
  H}$ loci in the figure are computed in the most conservative
approach: using the Lutz et al. relation, the dashed lines would be steeper
than those obtained from the Fiore et al. one. Therefore, the
intrinsic X--ray luminosity would be larger and, as a consequence, the
N$_{\rm H}$ needed to explain the observed luminosities of the MIRO targets would  also be larger.

\item
%\noindent
{\it X--ray spectra}:
The X--ray spectra of the three X--ray detected MIR/O are shown in
Fig.~\ref{MIROspectraXMM}.
Given the low photon statistics available for all the detected sources  
(in the range 20-150 net counts) we applied the Cstat statistic (\citealt{Cash1979}) to the unbinned data 
and assumed a very simple model in order 
to recover a rough estimate of the nuclear obscuration and instrinsic luminosity: 
a power-law with photon index fixed to $\Gamma=$ 1.9 plus obscuration at the source 
redshift (plus galactic N$_{\rm H}$).

In the XMM-detected sources, a second component is required by
the data in order to model the soft emission. 
Indeed, as demonstrated in \citet{Lanzuisi2015a}, given the
complexity of the X--ray emission, and the concurrent presence of
other processes such as scattering components or emission from star
formation, heavily obscured AGN can be missed when fitting
low-counting statistics data compatible with a single power-law model
(see also Lanzuisi et al. 2015b).
For all these sources we derived column density of N$_{\rm H}\approx$ 2.5-7$\times$ 10$^{23}$ cm$^{-2}$, with high uncertainties (see Table 2).  The X--ray detected sources, therefore, although characterized by a high obscuration, are not in the Compton Thick regime, as expected given current X--ray surveys limits and sensitivities and consistent with previous works \citep[e.g.,][]{Lanzuisi2009,Georgakakis2010}. The rest-frame  intrinsic X--ray luminosities are also listed in Table \ref{properties}. The most luminous of the three sources is MIRO20581 with an X--ray luminosity $\sim 10^{45}$ erg s$^{-1}$, while the other 2 sources have inferred intrinsic luminosities slightly below 10$^{44}$ erg s$^{-1}$. These luminosities exceed of about 2 dex those expected from stellar processes given the observed SFR (see below). \\

\item
%\noindent
{\it SED fitting}:
 Figure~\ref{SED} shows the SED fitting decomposition of the four targets considered in the paper, obtained making use of a modified version of the magphys code (\citealt{daCunha2008}) designed to take into account
a possible AGN emission component (\citealt{Berta2013}) together with a modestly-absorbed galaxy component. 
They have stellar mass M${_*}$ in the range 0.6-2$\times$10$^{11}$ M$_{\odot}$ and SFR of 10-100 M$_{\odot}$/yr, and are in (or below) the main sequence (MS) of star forming galaxies (see \citealt{Whitaker2012}), as suggested by their 'starbustiness' R$_{SB}$=sSFR/sSRF$_{MS,z}$ (see Table \ref{properties}). For two sources, MIRO20581 and MIRO28704, the far-infrared emission is not well constrained (see Figure~\ref{SED}), hence in the table are also reported SFR and R$_{SB}$ upper limits in parenthesis, computed treating the FIR upper limits as real detections. 

In all cases the (observer frame) NIR emission is dominated by the host galaxy light and the AGN disk component is considerably extincted (E(B-V)=2-4). According to the SED fitting decomposition, the 5.8$\mu$m luminosity is dominated by the torus emission ($\left \langle L_{5.8\mu m}^{torus}/L_{5.8\mu m}^{total} \right \rangle =87\%$), in agreement with the results from the bright DOG sample presented in \citet{Riguccini2015}. This also confirm that the X--ray-to-mid-IR diagnostic discussed above is a reliable instrument to test the X--ray obscuration of the MIRO targets.

According to the criteria exposed in \citet{Dey2008}  who classified dust obscured galaxies in AGN-dominated (``power-law'' DOGs) and SF-dominated (``bump DOGs'') on the basis of the rest-frame optical-to-mid-infrared SED shape (see their Section 3.1.2, and their Fig. 5), all the four MIRO targets would be classified as power-law DOGs, although MIRO18744 appear to show intermediate characteristic between the two classes.    
\item

\noindent
{\it Radio}:
All the targets are also detected in the  Very Large Array (VLA) observations of the COSMOS field (\citealt{Schinnerer2010}).  \citet{Fiore2009} reported that  QSOs  selected on the basis of the MIR/O excess at z$\sim$1.5 are more radio luminous than unobscured type 1 QSOs of similar luminosity and redshift, when the intrinsic 5.8 $\mu$m luminosities are compared (see also \citealt{Martinez2005}).  
MIRO20581 is the only radio loud target ($q_{24obs}=-0.6$, being $q_{24obs}=log(f_{24\mu
  m}/f_{1.4GHz})$; see \citealt{Bonzini2013}, Fig.2) in our SINFONI sample, with L$_{1.4GHz}$=4.5 $\times$ 10$^{25}$ W Hz$^{-1}$. 
MIRO20581 is also detected in the 3GHz survey of the COSMOS field
(Smolcic et al. 2015, in prep)  and is one of the most luminous
sources in the Very Long Baseline Array (VLBA) COSMOS catalogue (Herrera-Ruiz et al. 2015, in
prep); its Jansky Very Large Array (JVLA) measurements at 3 and 1.4GHz imply an inverted radio spectral index, which is consistent with a compact radio source rather than with a diffuse star forming region. Indeed, using the relation introduced by \citet{Condon1992}, between the 1.4 GHz luminosity and the SFR, we derive for  MIRO20581 a SFR a factor of 10-40 larger than the SED fitting estimate, and therefore the radio luminosity is interpreted as due to AGN activity. 
MIRO18744 and MIRO28704 are also detected in the VLBA catalogue and even in these cases, possible signatures of compact radio cores are present.

\item
%\noindent

{\it High ionisation diagnostics}:
For MIRO18744 a IMACS Magellan spectrum is available (\citealt{Trump2007}), sampling the rest-frame range 2850-4600 \AA. The spectrum shows a prominent [NeV]3425 emission line, an unambiguous sign of obscured nuclear activity (\citealt{Mignoli2013,Lanzuisi2015b}).

\end {enumerate}

The main properties of our SINFONI sample discussed above are reported in Table~\ref{properties}, with the targets sorted by decreasing MIPS flux and increasing redshift.
Overall, the multiwavelength constraints we have in our 24$\mu m$
bright  SINFONI
targets suggest unambiguously the presence of obscured AGN
activity. These characteristics correspond to those expected for
          objects caught in the post-merger dust-enshrouded phase
          of rapid black hole growth (see e.g. \citealt{Hopkins2008,Fiore2008,Fiore2009}).

\begin{figure*}
\centering
\includegraphics[width=7cm,angle=0]{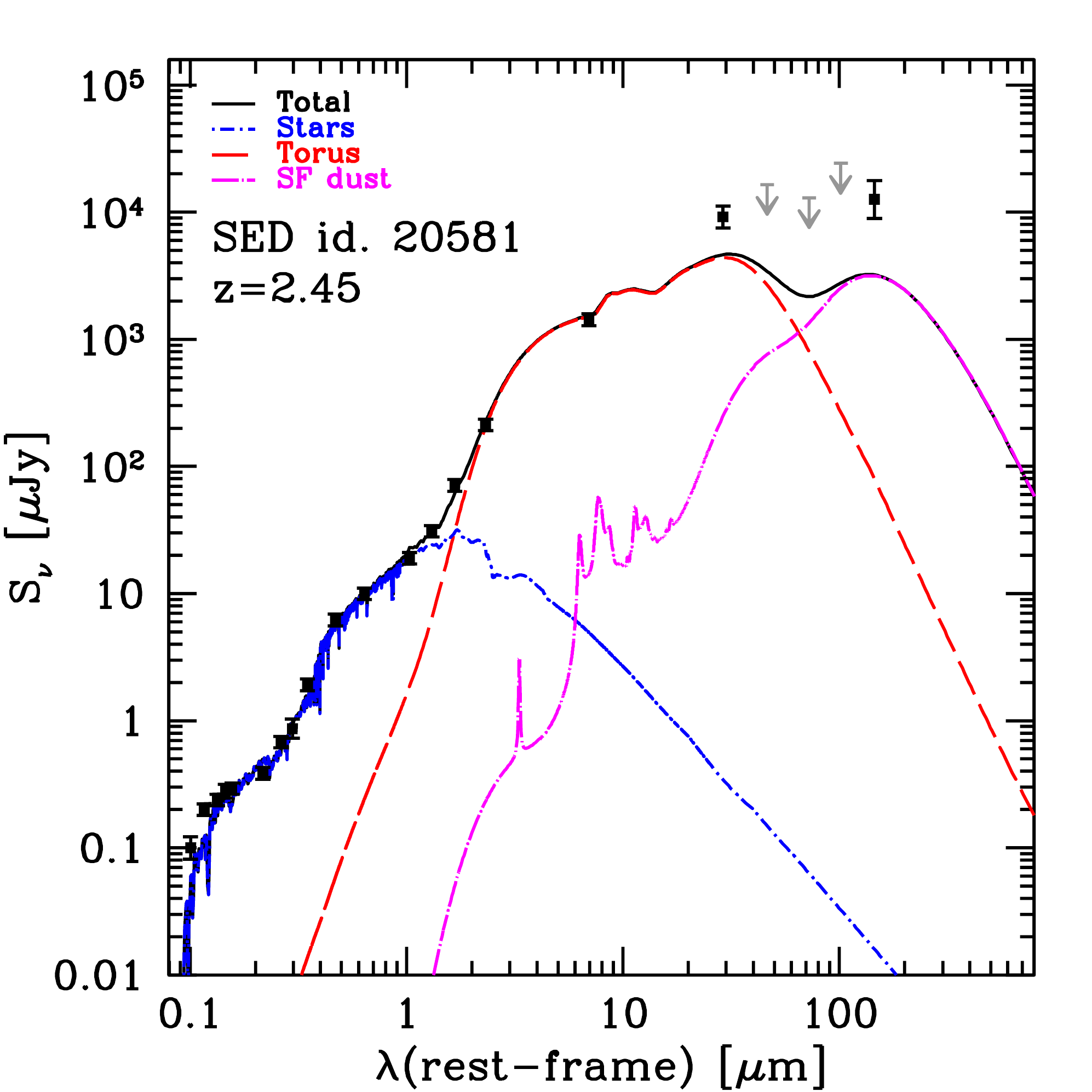}
\includegraphics[width=7cm,angle=0]{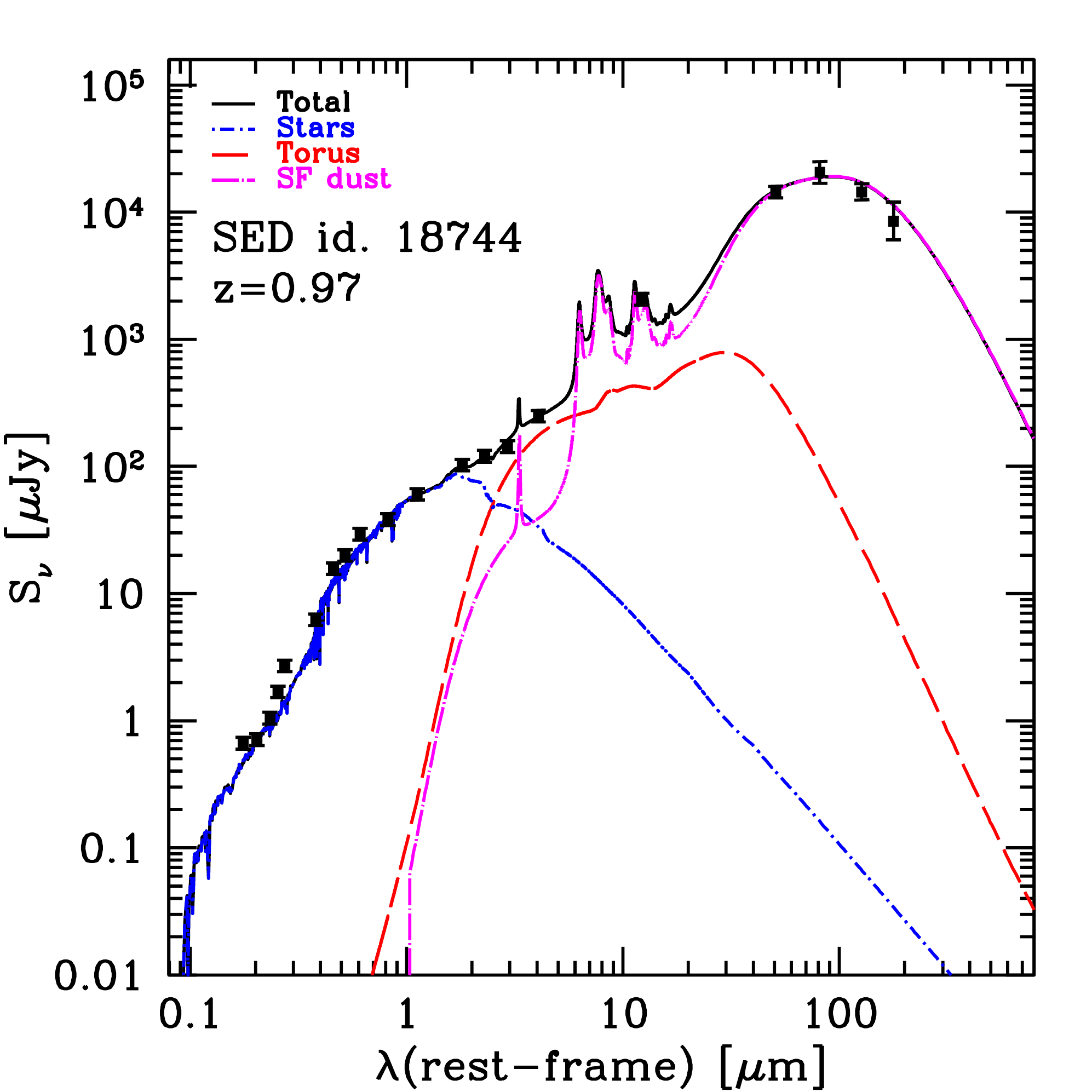}
\includegraphics[width=7cm,angle=0]{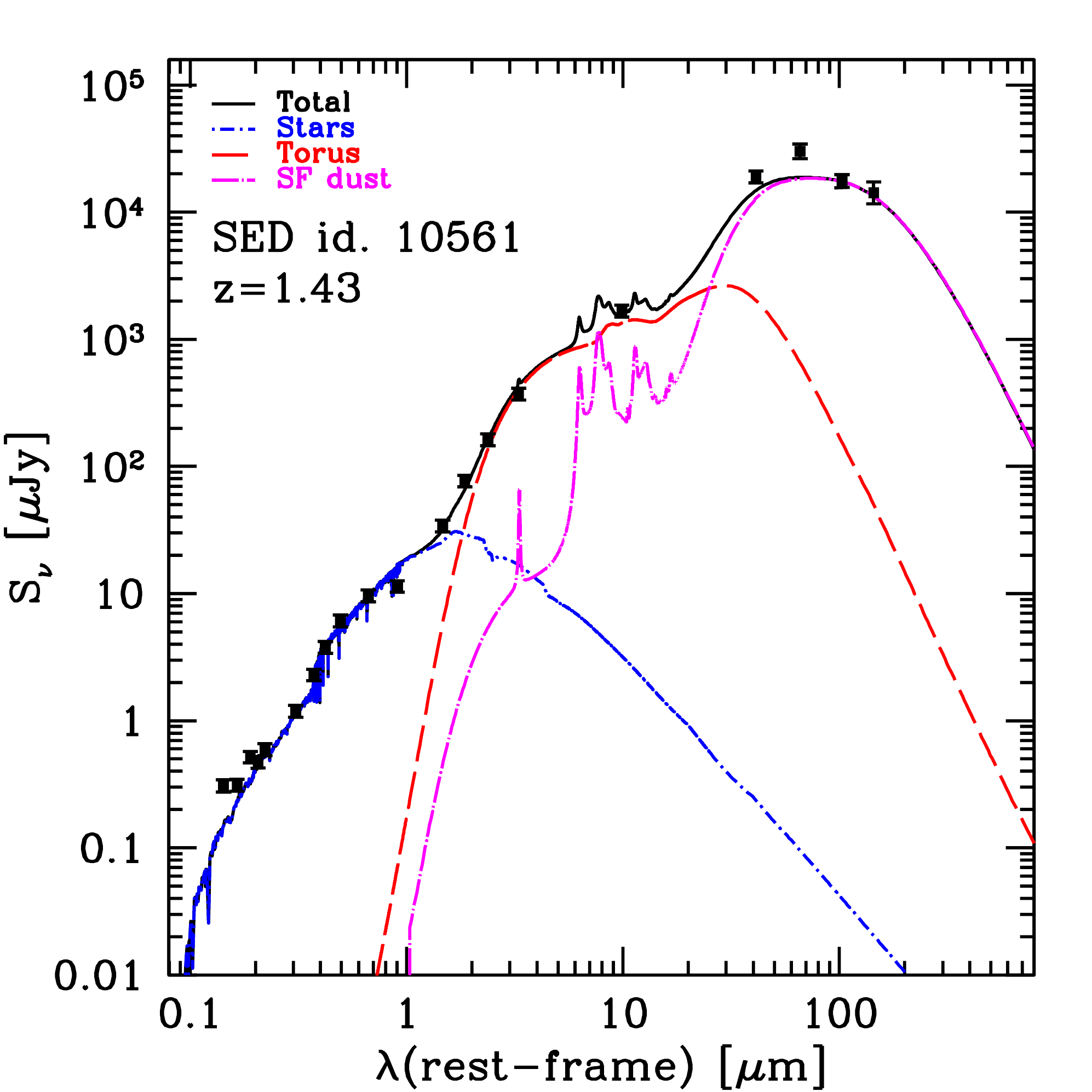}
\includegraphics[width=7cm,angle=0]{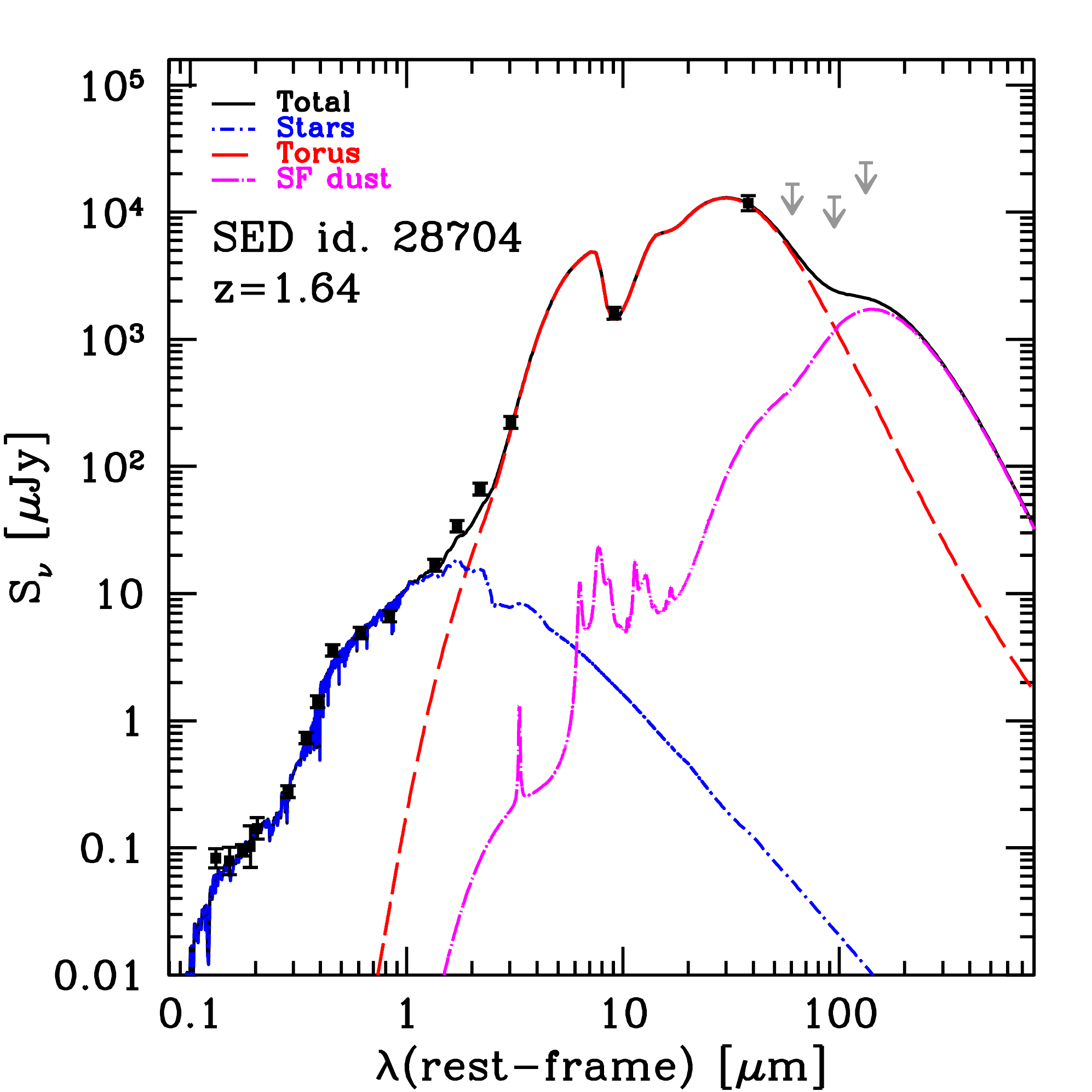}
\caption{SED fitting of MIRO targets. The black dots represent the
  observed data points ( in each panel, from left to right: SUBARU U, B, g, V, r, i, z; VISTA
    Y, J, H and Ks; the four Spitzer IRAC
  bands, Spitzer MIPS24; Hershel PACS and SPIRE). The blue line shows the integrated extincted emission originating from the host galaxy. The magenta line represents the star formation contribution for dust absorption, partially redistributed across the MIR/FIR range in a
self-consistent way (\citealt{daCunha2008,Berta2013, Delvecchio2014}). The red line reproduces the AGN contribution and incorporates both the accretion disc and the torus emission. The black solid line represents the sum of all components.}
\label{SED}
\end{figure*}

\section{SINFONI observations and data reduction}\label{sinfoniobs}

The observations were obtained in service mode using the near-infrared
spectrometer SINFONI of the VLT in adaptive-optics (AO)-assisted mode, during period 92A (from 2013-12-28 to 2014-03-30). All the targets in the sample were observed in one or two of the SINFONI filters (J, H, K or HK) 
depending on the initial redshift guess. We note that the program has been executed only partially ($\sim45$\%) and therefore our targets have not been observed with all the requested filters and/or for the entire requested time. We used a field-of-view (FoV) of 8x8'' in a 2D 64x64 spaxel frame. The spectral resolutions are R$\sim$1800 for J, R$\sim$2900 for H and R$\sim$1400 for HK.

We achieved a spatial resolution of 0.2" (FWHM) based on the  point spread functions (PSF) obtained in natural guide star (NGS) AO-mode, which roughly corresponds to 0.9 kpc at the average redshifts of z=1.5.  This spatial resolution is in agreement with those obtained in other SINFONI AO-assisted observations \citep[e.g.,][]{Bouche2013,Cresci2009}. 
Our targets do not extend more than $\sim$ 1-2'' in diameter, and were therefore observed with on-source dithering in order to use the object exposure with the closest MJD as an approximation of a sky exposure.
The information about the observations for each object are shown in Table~\ref{sample}. 

Besides the objects of the sample, a set of standard stars and their respective sky frames were also observed to flux-calibrate the data. Guide star names are also reported Table 1. The stars have R magnitudes in the range 15$<$R$<$16.

The data reduction process was performed using ESOREX (version 2.0.5). 
We used the IDL routine "skysub.pro'' (\citealt{Davies2007}) to remove the background sky emission. Then, we used our own IDL routines to perform the flux calibration and to reconstruct a final data cube for each object, adding the
different pointings. The flux calibration was performed following the
prescription indicated by \citet{Piqueras2012}. 

\section{Data analysis and Spectral fits}\label{fits}
%%%%%%%%%%%%%%%%%%%%%--------FIGURE START---------------%%%%%%%%%%%%%%%%%%
\begin{figure*}%[!h]
\centering
\captionsetup[subfigure]{labelformat=empty}
\begin{minipage}{1\linewidth}
\centering
\subfloat[a]{\label{20581}{ }\includegraphics[height=0.4\linewidth,width=0.4\linewidth]{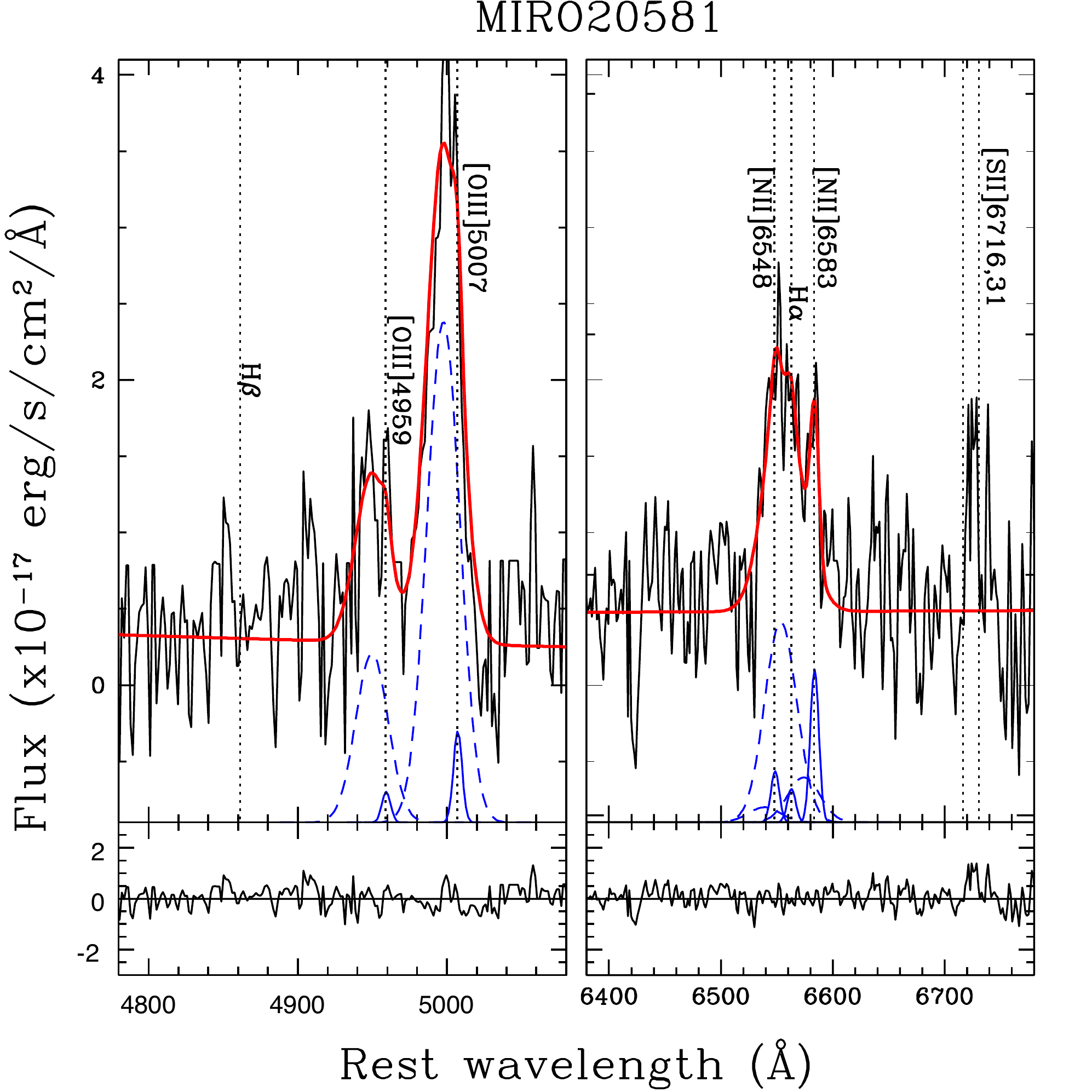}}
\subfloat[b]{\label{10561}{ }\includegraphics[width=0.4\linewidth]{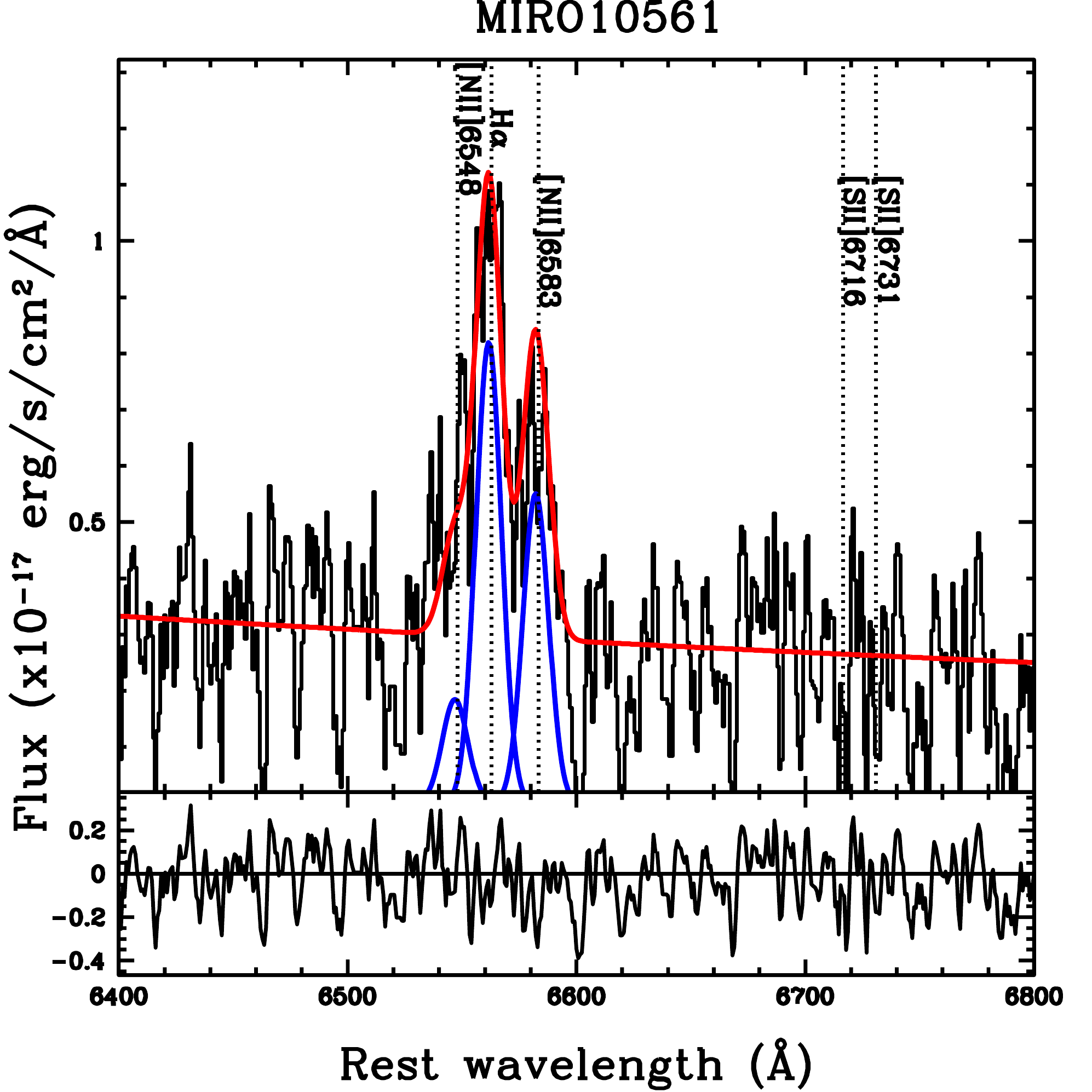}}
\end{minipage} \par\medskip
\begin{minipage}{1\linewidth}
\centering
\subfloat[c]{\label{18744}{ }\includegraphics[width=0.4\linewidth]{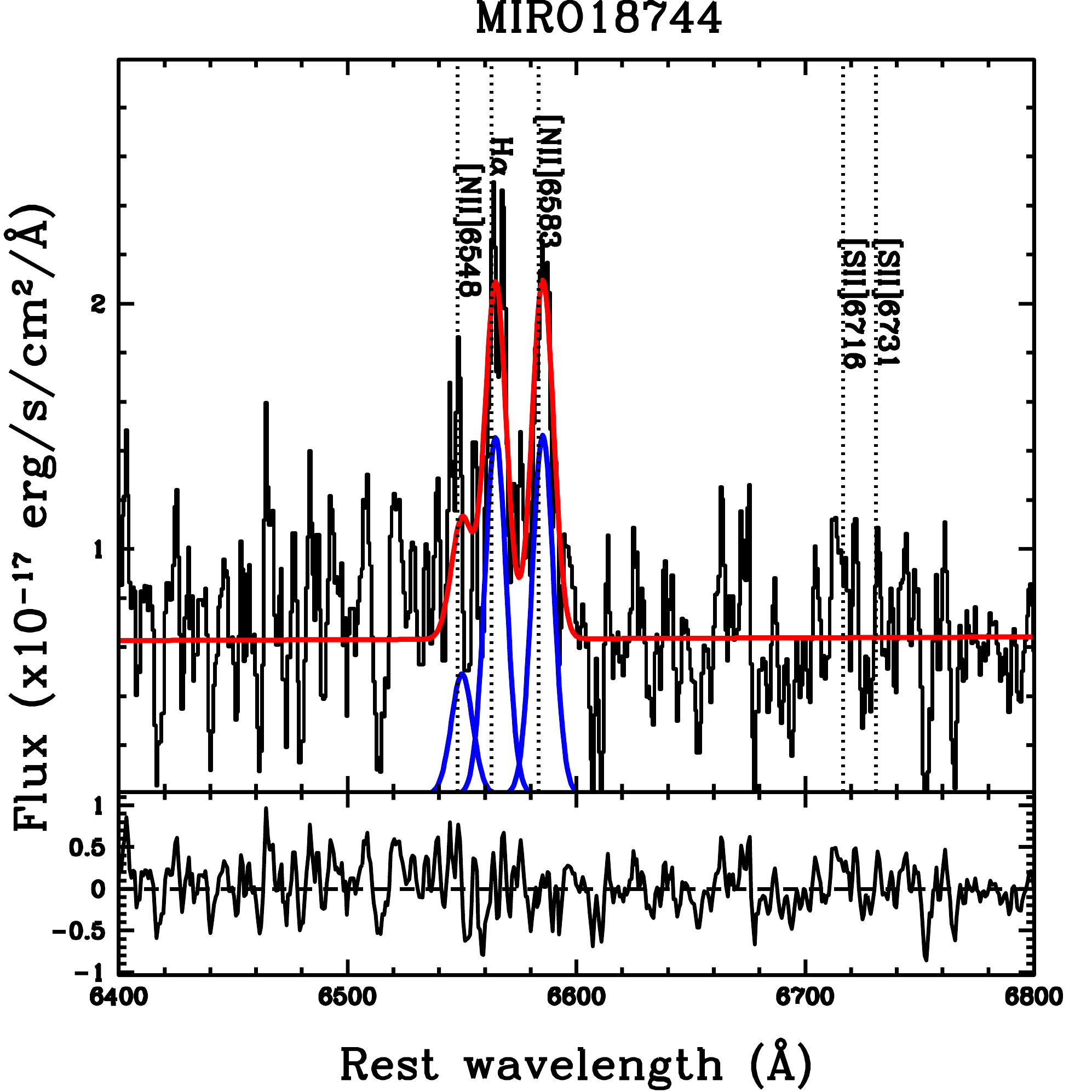}}
\subfloat[d]{\label{28704}{ }\includegraphics[width=0.4\linewidth]{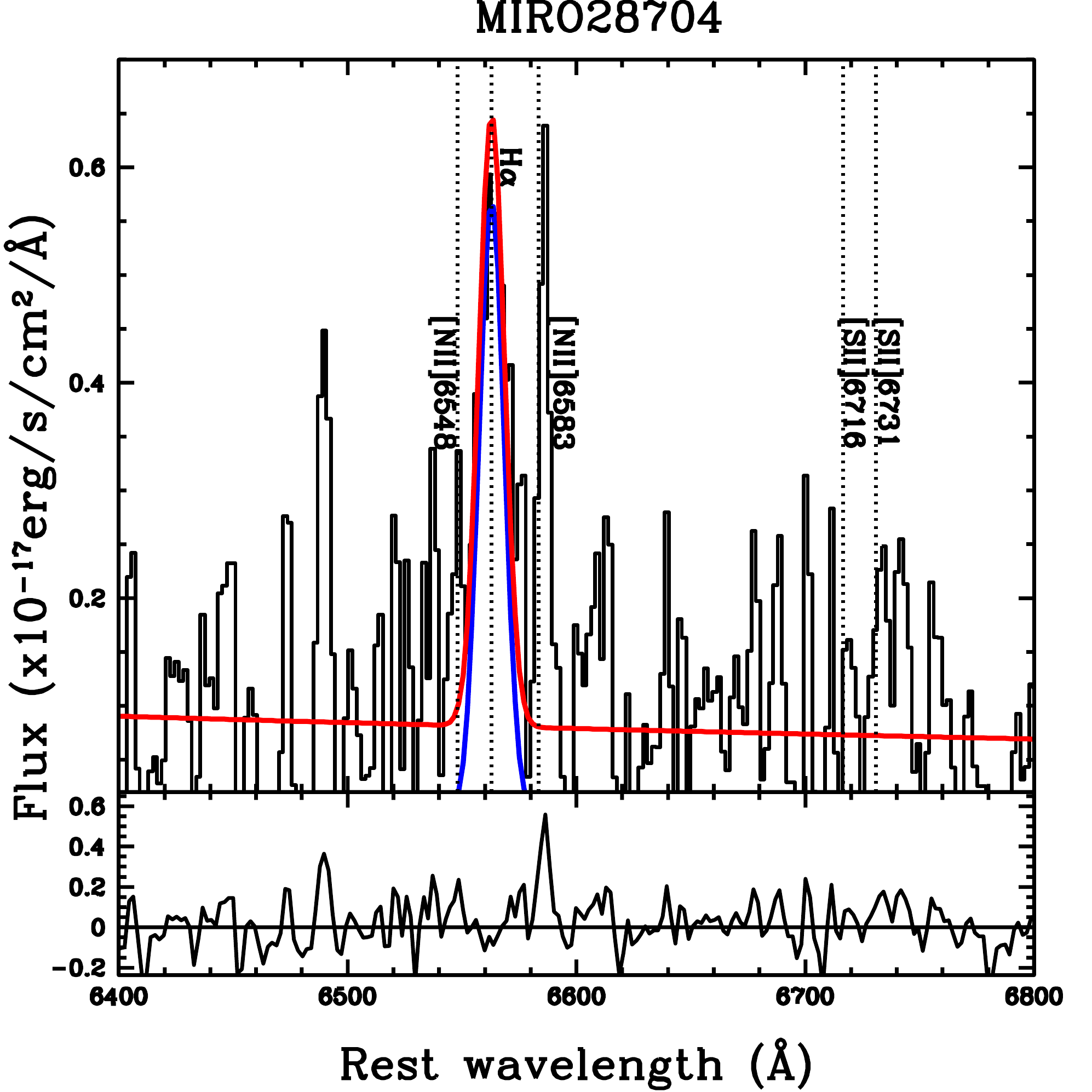}}
\end{minipage} \par\medskip
\caption{{\bf (a)} MIRO20581 HK band spectrum 
around the [OIII] (left) and the H$\alpha+$[NII] complex
(right). Superimposed on the spectrum are the best fit components
(solid and dashed blue curves, with arbitrary normalization in order
to ease the visualization). The red solid curves represent the sum of
all components, including the power-law. Dotted lines mark the
wavelengths of the H$\beta$, [OII] doublet, H$\alpha$, [NII] and [SII]
doublet.  {\bf (b)} MIRO10561 H band integrated spectrum around the H$\alpha+$[NII] complex.{\bf  (c)} MIRO18744 J band integrated spectrum around the H$\alpha+$[NII] complex.; {\bf (d)} MIRO28704 HK band integrated spectrum around the H$\alpha+$[NII] complex.  See (a) for the description of superimposed curves in (b), (c) and (d) panels. In the bottom panel of each fit the residuals with respect to the best fit are shown.}
\label{fig:spectraintegr}
\end{figure*}
%%%%%%%%%%%%%%%%%%%%%-----------FIGURE
%%%%%%%%%%%%%%%%%%%%%END------------------%%%%%%%%%%%%%%%%%%

Here, we briefly discuss the general data analysis and results of the spectral fits. In the next sections we describe the more detailed spatially-resolved analysis for our best-case, MIRO20581. 

Figure \ref{fig:spectraintegr} shows the one-dimensional integrated spectra, extracted in a 1-2'' diameter aperture, according to the compactness of the source. 
All the targets show the H$\alpha$+[NII] complex, with the exception
of MIRO28704 which has a lower-quality spectrum and in which only the
H$\alpha$\footnote {The narrow feature at $\lambda \approx$6584
$\AA$ is associated with a wrong skyline subtraction. We discard the
possibility that the observed line is HeI10830 at z=0.6. We fitted the
SED imposing this redshift. This fit produced a significantly larger
Chi-square, 4 times the value obtained imposing z=1.64. Indeed, at z=0.6
the source would be undetected down to a luminosity of log(L$_{X}$
)$\sim$ 42.2 in the 2-10 keV band; with a bolometric luminosity
estimated by this SED fitting decomposition of log(L$_{bol}$)=45.3,
MIRO28704 would have a very unusual bolometric correction of the order of
k$_{bol}$=L$_{bol}$/L$_{X} >$ 1000.} emission is detected. 
The only target for which we detected also the [OIII]5007 emission
line is MIRO20581. MIRO10561 was observed in the J band too, but both
[OIII] line and continuum emission were not detected. Instead, [OIII] emission
line for the other two targets was not covered, because of the
incompleteness of the observations (MIRO28704; see Section
\ref{sinfoniobs}) and of the wavelength coverage of SINFONI instrument
(MIRO18744; [OIII]5007 expected at $\lambda_{obs-frame} \approx$
0.986$\mu m$).

We fitted simultaneously each of the emission lines (from only one, H$\alpha$ in MIRO28704 to a total of six, H$\beta$, [OIII] doublet, [NII] doublet and H$\alpha$, for MIRO20581) with Gaussian line profiles.  When more than one emission line is fitted, we constrained the centroids and the line flux ratios according to atomic physics, while the widths were fixed to be the same as in each emission lines (see \citealt{P15}).

From the fit described above, we computed spectroscopic redshifts for our targets. We chose as redshift solution the one which produces the best fit of the narrow components of the emission lines.
We detected H$\alpha$ and [NII] emission for MIRO18744 at z=0.97, consistent with the spectroscopic redshift already available from the Magellan spectrum. For the remaining three sources, we were able to assign for the first time a spectroscopic redshift from our line fit (see column 13 in Table~\ref{sample}). The spectroscopic redshifts are in general agreement with the photometric estimates available within the COSMOS  survey  (\citealt{Salvato2011,Ilbert2009}), with an accuracy of $|z_{phot}-z_{spec}|/(1+z_{spec})\lesssim$ 0.1.

The results of the emission line fits are reported in Table \ref{spectralfit}.
In order to investigate the nature of the ionising source, we investigate the  emission line ratios diagnostics [NII]/H$\alpha$ and [OIII]/H$\beta$. The only source for which we have  both [NII]/H$\alpha$ and [OIII]/H$\beta$ (MIRO20581) to calculate the BPT diagram (\citealt{Baldwin1981}), lies in the AGN photoionisation region in such diagnostic.
For the remaining sources, although with large uncertainties due to the low-quality spectra, two out of three also show [NII]/H$\alpha$ ratios consistent with an AGN origin (\citealt{Veilleux1987}), consistent with the AGN classification discussed in Section 2.1. The diagnostic line ratios are reported in Table \ref{spectralfit}; lower limit in the [OIII]/H$\beta$ ratios are due to the non-detected H$\beta$ emission line.

We note here that although the clear presence of two objects in the
ACS-HST image of MIRO18744 (Figure \ref{cutouts}), 
given the low S/N, the SINFONI spectrum is relative to both objects.

\begin{figure}
\centering
\includegraphics[width=9cm,height=5cm,angle=0]{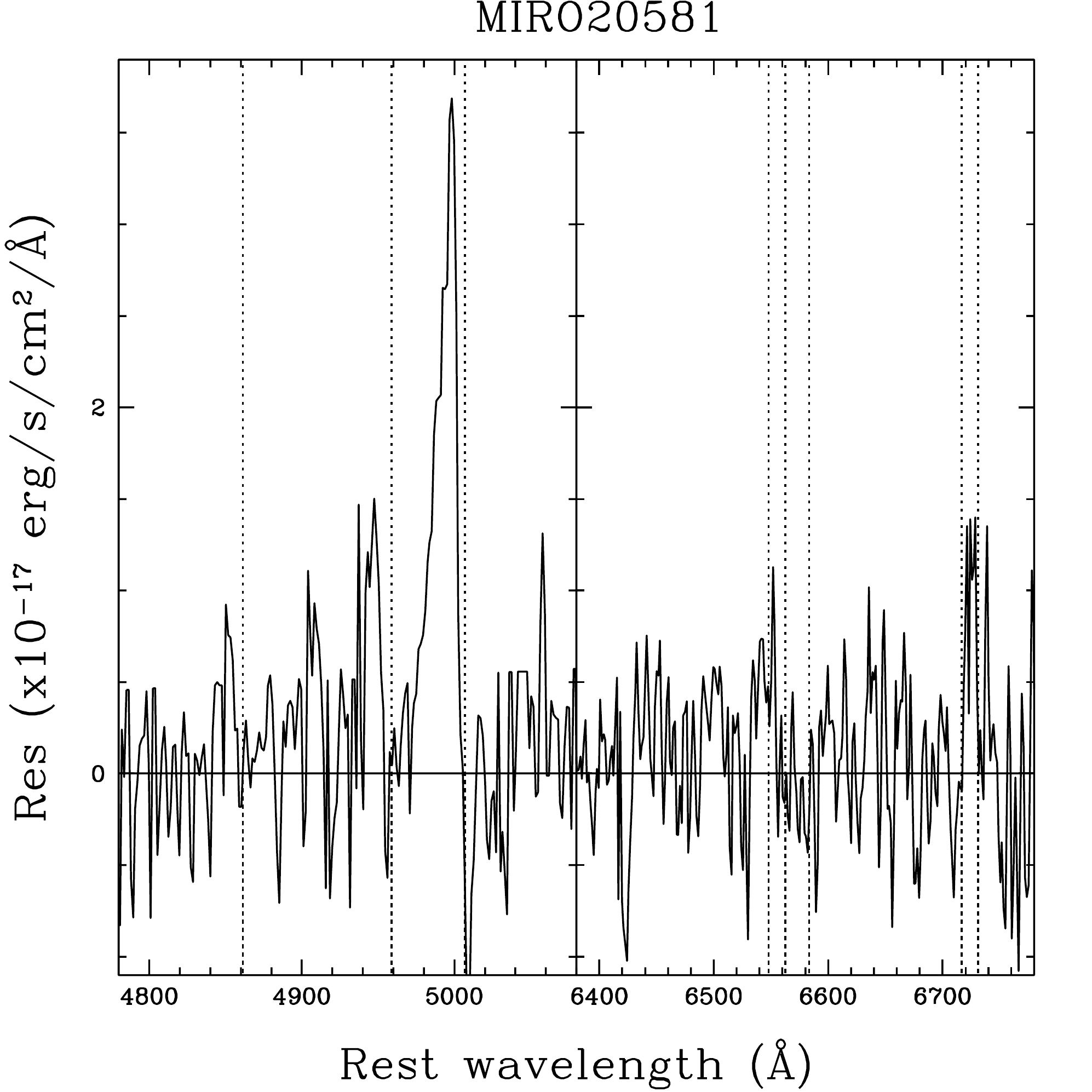}
\caption{MIRO20581 residuals in the [OIII] and Halpha range obtained by fitting the observed lines with only NLR and BLR components (See Section 5; to be compared to Panel a of Figure \ref{fig:spectraintegr}). }
\label{res20581}
\end{figure}

\begin{table*}
\footnotesize
\begin{minipage}[!h]{1\linewidth}
\centering
\caption{Emission line properties in the integrated spectra}
\begin{tabular}{lcccccc}
 MIRO  & &$f_{H\alpha}$    & $f_{[OIII]}$   & log($f_{[NII]}/f_{H\alpha}$) &log($f_{[OIII]}/H\beta$) & FWHM\\
       & & \multicolumn{2}{c}{(10$^{-17}$ erg/s/cm$^2$)}& & & (km/s)\\
(1)      & (2)& (3)   & (4)              & (5)  & (6) & (7)\\
\hline
18744&   &32.4$\pm$4.9& -& 0.12$\pm$0.08&-&565$\pm$20\\
10561&   &11.8$\pm$1.3& -& -0.10$\pm$0.07&-&560$\pm$25\\ %nii/Ha distribuzione simmetrica
28704&   &9.4$\pm$1.9 & -& -          &-&430$\pm$100\\
20581&NC    &4$_{-2}^{+3}$&2.5$_{-1.5}^{+2.5}$&0.88$_{-0.24}^{+0.30}$&$>-0.47$&360$\pm$70\\
     &OC&43.5$\pm$14.8&95$\pm$6&-0.16$\pm$0.51&$>0.47$&1600$\pm$100$^b$\\
     &NC$^a$&4$_{-2}^{+3}$&2.5$_{-1.5}^{+2.5}$&0.55$_{-0.65}^{+0.30}$&$>-0.47$&370$\pm75$\\
     &OC$^a$&24$\pm$13&95$\pm$6&-0.28$\pm$0.6&$>0.47$&1600$\pm$100$^b$\\
\hline
\end{tabular}
\label{spectralfit}
\end{minipage}
Notes: (1) target name; (2) kinemetic component: NC = narrow
  component, OC = outflow component; (3) H$\alpha$ flux;
(4) [OIII]5007 flux; (5) diagnostic [NII]6583/H$\alpha$ flux ratio;
(6) diagnostic [OIII]5007/H$\beta$ flux ratio; (7) kinematic component
width. \\
$^a$ 
These results were obtained  following the second approach illustrated in Sec. \ref{sec20581fit}, adding a BLR component for the H$\alpha$ profile.\\
$^b$ The centroid of the OC profile is blueshifted of $\sim$ 550 km/s.

\end{table*}

\section{The ionised outflow in MIRO20581}\label{sec20581fit}
 We now concentrate on MIRO20581.
In order to reproduce the line profiles in the [OIII] and H$\alpha$
region we had to introduce an extra blueshifted and broad
(FWHM$\approx1600$ km/s) component (see Figure
\ref{fig:spectraintegr}, panel $a$). Emission from ionised gas in
forbidden lines like [OIII] can not be associated to motion in the
broad line region (BLR) because it would be suppressed by collisional de-excitation when produced in high-density regions (but see \citealt{Devereux2011} for an alternative explanation). For this reason, any broad ($>550$ km/s) profile in such forbidden lines is generally interpreted as to be ascribed to outflowing ionised gas. 
Broad H$\alpha$ profiles may be also ascribed to the presence of a
ionised outflowing gas (e.g. \citealt{Genzel2014}); however, the
H$\alpha$ emission may suffer from severe contamination by the
presence of the BLR motion and therefore may be considered less
reliable tracer of outflows and the associated energetics. This is
especially the case when the high-velocity BLR wings  are not detected, as instead, for instance, it is the case of Mrk 231 (\citealt{Rupke2011}).

In order to investigate at best the ionised gas emission in MIRO20581,
and taking into account the considerations above, we fitted
simultaneously the H$\beta$+[OIII] and H$\alpha$+[NII] regions (see
Section \ref{fits} for details) using
two approaches. In order to reproduce the line profiles of all the
emission lines, we fitted the two regions
\begin{enumerate}
\item 
with two sets of Gaussian profiles: one to account for the presence of
NLR components (in the following narrow component, NC), with FWHM$\lesssim$550 km/s, one for the  presence of outflow components (OC), with FWHM$>$550 km/s;
\item
using the same components as above, namely the NC and OC components, and adding a broader profile ( FWHM$>$1900 km/s) to account for the presence of the H$\alpha$ emission originated in the BLR.
\end{enumerate}

The two fits have acceptable low residuals and adequately represent the shape of the line profiles, hence it is not possible to confirm or exclude the presence of the BLR emission in this source, at least with this S/N. The results of these two approaches are reported in Table \ref{spectralfit}. If present, the H$\alpha$  BLR emission would have a FWHM of about 2500-3000 km/s. The lower value has been obtained fitting the entire H$\alpha$+[NII] profile with only one Gaussian. The NII/H$\alpha$ ratio depends on the detailed modeling but, in both cases, the emission lines remain in the AGN photo-ionisation region (see previous Section).
  
A tentative fit with 2 sets of Gaussian profiles, taking into account
only NLR and BLR components was performed. This has shown high
residuals especially in the [OIII]5007 profile. Figure
\ref{res20581} shows these residuals.

\subsection{Spatial analysis}
Figure \ref{cubo}, panel $a$, shows the contour plot of the median SINFONI data-cube over the entire HK wavelength range, in steps of 1$\sigma$ starting from 3$\sigma$ (green scale). The standard deviation $\sigma$ was computed  in a 1.25--1.75'' annulus centred on the target. The astrometry in the SINFONI data-cube was performed using the Ks COSMOS cutout, obtaining a match between the coordinates of the peak of intensity in the Ks COSMOS cutout and in the median SINFONI data-cube. In Figure 7a, the 
Ks COSMOS contours are over-imposed on the SINFONI data-cube (cyan solid curves). We also show the HST/ACS F814W contours at higher resolution (magenta curves; see also Figure 1).

Figure \ref{cubo}, panel b, shows the [OIII]5007 map integrated on the continuum-subtracted total line profile (grey scale), with over-imposed contour levels (starting from 3 $\sigma$) of the emission of the line core (5000-5014$\AA$; width $\sim800$ km/s) and of the blueward part (4973-5000$\AA$; width$\sim1600$ km/s) of the line profile. 
Blueshifted emission is found out to a distance of R=0.6'' (associated to region B), i.e. 4.8 kpc from the nucleus. Instead, the core emission is less extended, and could be associated with the narrow component of the [OIII] profile.  
\subsection{Kinematic analysis}\label{kinanalysis}
In order to map the line emission distributions and the corresponding velocities, 
a nuclear and a off-nuclear spectra are extracted from two 3x2 spaxel regions (3x2 kpc) close to the central and the off-nuclear peaks in the [OIII] channel map (Fig. \ref{cubo} panel $b$, regions A and B labelled with black boxes; see also \citealt{P15,Cresci2015}). 
In Figure \ref{cubo}, panel (c), we show the integrated spectra over the two regions: red and orange solid lines represent the nuclear and off-nuclear line profile respectively. 

In both cases, the [OIII] profiles are broad, with widths of FWHM=1400-1200 km/s, and maximum velocities of 1600-1650 km/s  (nuclear-offnuclear spectrum). In the off-nuclear region, the centroid of the emission lines is blueshifted of $\sim$700 km/s.

\subsection{Outflow properties}\label{outflowprop}
\begin{figure*}
%\begin{minipage}[t]{12.0cm}
\centering
\includegraphics[width=18cm,angle=0]{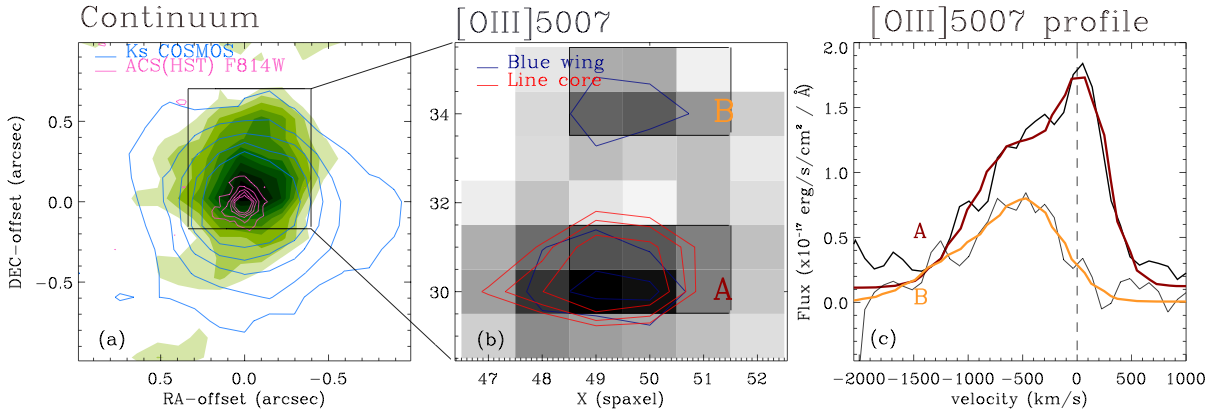}
%\captionsetup{font={small}}
%\end{minipage}
\caption{{\it (a)}: MIRO20581 contour plot of the median SINFONI data-cube over the entire HK wavelength range (green scale,  in steps of 1$\sigma$ starting from 3$\sigma$), with over-imposed the Ks COSMOS (cyan; starting from 3$\sigma$, in steps of 3$\sigma$) and ACS HST contours (magenta; starting from 3$\sigma$, in steps of 3$\sigma$). {\it (b)}: [OIII]5007 channel map obtained integrating the continuum-subtracted SINFONI data-cube on the total line profile (4973-5024$\AA$), of the region selected in panel (a).  The contours levels are in steps of 1$\sigma$ (starting from 3$\sigma$) and show the [OIII] emission coming from the core (5000-5014$\AA$) and from the bluer (4973-5000$\AA$) part of the line profile. {\it (c)}: [OIII] line profiles obtained from integrated spectra over the spaxels selected in the two regions ''A'' and ''B'' in the panel (b). Vertical dashed lines represent the systemic velocity obtained from the 1'' integrated spectrum. 
}
\label{cubo}
\end{figure*}

Assuming that the broad and shifted [OIII] component can be associated with an outflowing wind, the
kinetic power ($P_K^{ion}$) and mass-outflow rate ($\dot M_{out}^{ion}$) of the outflow can be computed under reasonable assumptions in the case of a biconical geometry. 
First of all, given that the electron density of the outflowing gas can not be estimated directly from the data due to the low quality of the spectra in the [SII] region (see Figure \ref{fig:spectraintegr}), and the metallicity indicators (\citealt{Pettini2004,Yin2007}) are not useful because of the AGN ionising radiation, we assumed standard values of $n_e$ (100 cm$^{-3}$) and metallicity (solar).
Generally, the H$\beta$ luminosity is used to calculate the total
amount of gas and the mass outflow rate
\citep[e.g.,][]{Liu2013,Cresci2015}. However, in our case this line is
not detected, and  we have adopted the \citet{CanoDiaz2012} formulae
which instead employ the [OIII] line luminosity:

\begin{equation}\label{canodiaz}
P_K^{ion}=5.17\times 10^{43} \frac{CL_{44}([OIII])v_{out,3}^3}{n_{e3} R_{kpc} 10^{\left [ O/H \right ]}} erg\ s^{-1},
\end{equation}
\begin{equation}\label{Mcanodiaz}
\dot M_{out}^{ion}=164\times 10^{43} \frac{CL_{44}([OIII])v_{out,3}}{n_{e3} R_{kpc} 10^{\left [ O/H \right ]}} M_\odot\ s^{-1},
\end{equation}

where $L_{44}([OIII])$ is the [OIII] luminosity associated to the outflow component in units of 10$^{44}$ erg s$^{-1}$, $n_{e3}$ is the electron density in units of 1000 cm$^{-3}$, $v_{out,3}$ is the outflow velocity $v_{out}$ in unit of 1000 km s$^{-1}$, $C$ is the condensation factor ($\approx$ 1), 10$^{[O/H]}$ is the metallicity in solar units, $R_{kpc}$ is the radius of the outflowing region in units of kpc. 
 We therefore used the [OIII]5007 flux associated to the outflow component in the 1'' integrated spectrum (Figure \ref{fig:spectraintegr}, panel $a$). We further adopted a spatial  extension of 4.8 kpc for the outflowing gas  given that we observe the blueward emission out to this distance (see Figure \ref{cubo}, panel $b$; section 5.1).
Finally, we considered as outflow velocity the maximum velocity observed $v_{max}$ in the nuclear region ($v_{out}=1600$ km/s; see Section \ref{kinanalysis}), and we assumed that lower velocities are due to projection effects (\citealt{CanoDiaz2012, Cresci2015}).

Following the Cano-D\'\i az et al. formalism, the kinetic power is
$P_k^{ion}$=1.5$\times$ 10$^{44}$ erg s$^{-1}$, while the outflow mass
rate is $\dot M_{out}^{ion}$=190 M$_\odot$ yr$^{-1}$, consistent with the values observed for targets at similar bolometric luminosities  \citep{Carniani2015}. These equations assume a simplified model where the wind occurs in a conical region uniformly filled with outflowing ionised clouds. The values, not corrected for the extinction and regarding only the ionised component of the outflow (but see also the other conservative conditions in \citealt{CanoDiaz2012}, and the discussion in \citealt{P15}, section 6.1), represent lower limits to the total outflow power. 

The kinetic power is $\approx 0.4\%$ of the AGN bolometric luminosity
also inferred from the SED fitting decomposition, in rough agreement
with the predictions of AGN feedback models (few \% of $L_{bol}$,
\citealt{King2005}\footnote{ We note however that our kinetic power
  estimate is
  related to the only ionised component, while the prediction interest
  all the outflow components (i.e., molecular, atomic and ionised components).
}). 
The momentum flux,  $\dot P_k^{\ ion}$=$\dot M v_{out}$ is
2$\times$10$^{36}$ dyne, $\approx$ 2 times the radiative momentum flux
from the central black hole, $L_{bol}/c$. Its momentum boost, i.e. the
ratio between $\dot P_k^{ion}$ and $L_{bol}/c$ is in agreement with
those observed in X--ray winds \citep[e.g.,][]{Tombesi2015} rather
than with the ratios associated with ionised and/or molecular outflows
\citep[e.g.,][]{Cicone2014,P15}. This discrepancy however, could be
totally attributed to the fact that our estimate of the momentum flux
represents a lower limit. In fact, correcting the [OIII] luminosity
for an E(B-V)$=$1.3 calculated from the SED fitting
decomposition\footnote{This value has been obtained  as rough estimate
  assuming for the outflowing material a reddening value in between
  the E(B-V) estimated for the AGN and galaxy components (see Table
  \ref{properties}). In fact, possible differential obscuration
  between nuclear and off-nuclear regions may be present.}, we obtain
a mass outflow rate a factor of $\sim10$ larger ($\sim2000$
$M_\odot/yr$) and, consequently, momentum flux and
momentum boost in a more reasonable agreement with the results
reported in the literature (namely: a momentum boost 20 times
$L_{bol}/c$, a kinetic power $\approx 4\%$ $L_{bol}/c$), and favouring an ``energy-conserving'' nature of the observed large-scale ionised outflow (see e.g. \citealt{Carniani2015} and references therein). \\
However, these results are based on few assumptions whose contribution
can be remarkable. Estimates of the electron density of outflowing
region have been obtained in few low-z AGN and Ultra-Luminous Infrared
Galaxies  \citep[e.g.,][]{Villar2014, Zaurin2013} and high-z
QSO  (e.g., \citealt{P15}, Brusa et al. in prep), with values between 10$^2$ and 10$^3$
cm$^{-3}$. Although the value we used in Eq. \ref{canodiaz} and
Eq. \ref{Mcanodiaz} is the one routinely used in
literature  \citep[e.g.,][]{Liu2013,Harrison2014,Cresci2015}, outflow
energetics may be a factor of 10 lower if n$_e=$10$^3$ cm$^{-3}$ is adopted. A further factor of 1/2 should
be considered, taking into account
the assumption on the metallicity, if metal-rich regions are present
(see \citealt{P15}). Finally, we considered a reddening value
in between the E(B-V) estimated for the AGN and galaxy components (see Table
  \ref{properties}). Considering the AGN (galaxy) reddening E(B-V) estimate instead of the average, 
  the [OIII]5007 flux and therefore all the energetics should be
  corrected of a factor of 100 (5). 
We
  note however that all basic assumptions previously outlined are in agreement with
  those adopted in similar studies in literature and, in
  the case of the reddening value, reasonable. For completeness, we
  report in Table \ref{energetics} all the
  energetic values with ranges obtained minimizing and maximizing the Eq. \ref{canodiaz} and
Eq. \ref{Mcanodiaz} using all the possible alternative assumptions above described.

\begin{table}
\footnotesize
\begin{minipage}[!h]{1\linewidth}
\centering
\renewcommand{\arraystretch}{1.2}
\caption{MIRO20581 outflow energetics}
\begin{tabular}{l|ccc}
  & basic asm & min. asm & max. asm\\
    & (1)   & (2)  & (3)\\ 
\hline
$\dot P_k^{\ ion}$ (erg s$^{-1}$) & 1.5$\times$10$^{45}$ & 3.8$\times$10$^{43}$ & 4.6$\times$10$^{46}$\\
\hline
$\dot M_{out}^{ion}$ (M$_\odot$ yr$^{-1}$) & 2000 & 48 & 20000\\ 
\hline
$P_k^{\ ion}/L_{bol}$ & 0.04 & 0.001 & 0.4\\
\hline
$\dot P_k^{\ ion}/(L_{bol}/c)$ & 20 & 0.5 & 200\\
\hline 
\end{tabular}
\label{energetics}
\end{minipage}
Notes: Energetic values obtained from Eq. \ref{canodiaz} and
Eq. \ref{Mcanodiaz} using: basic assumptions (column 1), 
minimizing assumptions (column 2) and maximizing assumptions (column
3) on n$_e$, metallicity and extinction.
\end{table}

\section{Discussion}\label{discussion}
We analyzed the NIR SINFONI spectra of four candidate obscured QSOs, selected from the COSMOS survey, on the basis of red mid-infrared-to-optical and optical-to-near-infrared colours.

Broad profiles in the [OIII] and H$\alpha$ lines with FHWM $>550$ km/s (OC component) which are commonly used as signposts of outflows have been detected  only in one source, MIRO20581.
 We can not exclude the presence of faint OC components in the other sources, given the low quality of the spectra and the lack of the [OIII]5007 emission, that is a better optical tracer for outflows. 
Overall, the integrated spectra of the other three sources have low S/N  (with emission lines detected at $\sim$ 3 $\sigma$), and faint outflow components in the H$\alpha$+[NII] could still be present. 
Alternatively, the high obscuration of the sample, proofed the SED
decomposition, the X--ray spectra, and the X--ray-to-mid-IR ratios
(see Section \ref{multiwav}), might suggest that they are instead in
the rapid black hole growth phase, when the scaling relations between
host galaxies and black hole properties are not yet established and
winds have not been launched yet (\citealt{King2005}). If it is
  the case,
MIRO20581 is different from these other three sources. We investigate
this scenario in the following section.

\subsection{Comparison with literature}
In this section we compare the properties of our MIRO targets with those of a sample of AGN-dominated DOGs reported in the literature, and discuss their main similarities/differences. 
Several studies have shown that DOGs with large $f_{24\mu m}$ flux ($\gtrsim1$ mJy) exhibit higher AGN activity, higher concentration and smaller physical size. Vice versa, DOGs with lower $f_{24\mu m}$ exhibit higher SF activity and larger physical size (\citealt{Melbourne2011,Riguccini2015}). Although it is shown that the rest-frame optical morphologies of the most luminous DOGs have little sign of ongoing mergers (\citealt{Melbourne2008,Melbourne2009,Bussmann2009}), there are also indications of non-regular gas kinematics in their host galaxies (\citealt{Melbourne2011}) and, moreover, several arguments suggest that they could be post-merger products of gas-rich mergers (see e.g. {\citealt{Melbourne2009}). Merger simulations were able to reproduce their colours and luminosities and indicate an infrared-to-optical drop as gas consumption and AGN-driven wind terminate both SF and BH growth (\citealt{Narayanan2010}).

\citet{Brand2007} studied NIR Keck spectra of a sample of ten AGN-dominated DOGs selected in the
 9 deg$^2$ NOAO Deep Wide-Field Survey Bo\"otes field (\citealt{Jannuzi1999}). Most of their spectra have low S/N and it is not possible to rule out the presence of outflows in the observed emission lines, but a prominent broad (FWHM $\sim$ 1600 km/s) [OIII]5007 profile was detected in one target, SST24 J1428+34. H$\alpha$ or H$\beta$ BLR have been found in 70\% of them. 
\citet{Melbourne2011}, instead presented a sample of four AGN-dominated DOGs with high spatial resolution Keck OSIRIS integral field spectroscopy.  The sources were selected also in the 
 Bo\"otes field, and the main selection criteria was the strong H$\alpha$ BLR  detection in available NIR spectroscopic observations (e.g., from the same \citealt{Brand2007} sample). They found that the BH masses of their sample are small for their host galaxy luminosities when compared with z $\sim$ 2 and local unobscured AGNs (see their Section 5.1). Indeed, they did not find any evidence of outflows in the hosts, accordingly with the above-cited predictions of \citet{King2005}. They also reported SFRs $<100$ $M_{\odot}\ yr^{-1}$ for all targets.
Finally, the \citet{B15} sample has been selected on the basis of red R-K colours (with a cut at K$_{AB} <19$), and high X/O ratio. The latter selection criterion is roughly equivalent to the high MIR/O ratio (see \citealt{Fiore2008}). To confirm this we added the two brightest sources (XID2028 and XID5321; \citealt{P15}) in the \citet{B15} sample 
in Figure \ref{mirofiore} (with blue star and square), as representative of the entire sample.
Evidence of outflows have been found in 75\% of object, and BLR emission have been found with similar percentage. 

Overall, all these sources have similar colours and MIR/O ratios to those of our MIRO targets (see Figure \ref{mirofiore}; cyan stars and squares represent the sources in \citet{Melbourne2011} and \citet{Brand2007} samples with K-band measurement).
While BLR emission may be present only in 1 out 4 of our targets (MIRO20581; see Section \ref{sec20581fit}),
a large number of \citet{B15}, \citet{Melbourne2011} and \citet{Brand2007} obscured QSOs exhibit BLR emission. 
However, as already mentioned, the  \citet{Melbourne2011} targets were
pre-selected to have a strong H$\alpha$ detection and, in general, all
the targets in these three comparative samples have strong K-band
emission (K$_{AB} \lesssim 20$): the \citet{B15} sample has been
selected requiring a K$_{AB} <19$;  the Bo\"otes field from which the
\citet{Melbourne2011} and \citet{Brand2007} DOGs were selected, has
considerably shallower NIR observations than the COSMOS field (K$_{AB}
< 23$, \citealt{McCracken2010}; K$_{AB} < 20.8$, \citealt{Dey2008},
respectively). On the contrary, for the observations proposed in this
work, we did not impose any flux threshold. Hence, it seems that the
presence of BLR emission is related to the K-band flux. This offers a
possible interpretation. All these targets are AGN-dominated DOGs at
z$\sim2$; therefore, an higher K-band flux may correspond to an higher
rest-frame optical AGN continuum emission, that is proper of AGN
showing BLR emission. Vice versa our sample, with K$_{AB}\gtrsim 20$,
is dominated by rest-frame optical host-galaxy continuum emission (see
Figure \ref{SED}). The X--ray absorption is fully consistent with type
2 (1.9 in the case of MIRO20581) classification obtained by the SINFONI spectra.

The only object in which we detected the outflow, MIRO201581, stands out with respect to the other targets. 
Although sharing the same X--ray luminosities of the \citet{B15} targets, the X--ray spectrum of MIRO20581 shows high column density (N$_H\approx$ 7 $\times$ 10$^{23}$ cm$^{-2}$), 
larger than those observed for the X/O targets (N$_H\approx$10$^{21.6}$ cm$^{-2}$, \citealt{P15}). This difference may be attributed mainly to line of sights effects which intersect a larger portion of the torus in MIRO20581 with respect to the \citet{P15} targets, and would also explain the fact that in MIRO20581 we do not detect a dominant BLR component in the H$\alpha$. Alternatively, the high extinction seen in the SED in both the AGN and host-galaxy components may be related to large scales obscurations. 

Assuming that the H$\alpha$+[NII] complex of MIRO20581 is composed of NC, OC  and a BLR component, its black hole mass can be estimated using the \citet{Bongiorno2014} formula, assuming a FWHM of the H$\alpha$ BLR emission of 3000 km/s (see Section \ref{sec20581fit}), log(M$_{BH}$/M$_\odot$)=8.4. Considering the stellar mass estimated by SED fitting decomposition (log(M$_*$/M$_{\odot}$)=11), we measure a central black hole to stellar mass ratio of the host of
$\sim$0.002, comparable to the ratios of unobscured QSO at the same
redshift (\citealt{Bongiorno2014,Merloni2010}).  Hence, it seems that outflows are present only in sources for which the $M_{BH}-M_{*}$ relation has been fixed, in agreement with the predictions of  \citet{King2005}.
 
From the comparison between L$_{bol}$ (Table \ref{properties}) and the Eddington luminosity associated to the M$_{BH}$, we also infer an high Eddington ratio (L$_{bol}/L_{Edd}\sim1$). All these properties (high Eddington ratio, high extinction, M$_{BH}$/M$_*\sim0.002$) point towards the interpretation that MIRO20581 could be associated with the beginning of the blow-out phase.
On the other hand, its SFR and starbustiness are low (but still within $\pm$0.6 dex of the MS; see Table \ref{properties}), indicating perhaps an advanced state of the ongoing process of negative feedback (see also \citealt{Balmaverde2015}). 
We underline that its FIR emission is not well constrained (see Figure
\ref{SED}); hence we calculated as upper limit a SFR $=132$
M$_\odot$/yr, treating the FIR upper limits as real detections. 
An higher SFR, and consequently an higher sSFR, may be more consistent with the beginning of a blow-out phase, in which the effects of feedback are still marginal and the SF is still occurring. 

In the \citet{Narayanan2010} simulations, AGN-dominated DOGs appear after the peak of the star formation, and therefore not necessarily associated with Starburst phase, given that the time scales of the two processes are not the same. 
Indeed, results recently presented in \citet{Riguccini2015} showed
that 50\% of AGN-dominated DOGs detected by Herschel display sSFRs that place them in or above the MS, while the remaining 50\% are below the MS, indicating perphaps an ongoing quenching of the star formation due to the AGN activity.  
In this scenario,  small SFRs observed in dust-obscured massive main
sequence galaxies hosting AGNs such those of MIRO10561, MIRO28704,
MIRO18744 and those presented in \citet{Melbourne2011} may be associated with systems which are still actively growing their black holes.

\section{Summary}
The main results from the multiwavelength analysis and the SINFONI data on a small sample of mid infraread bright, red quasars presented in this paper are summarised below:
\begin{itemize}
\item
 All the sources but MIRO28704, selected from the 24$\mu m$ Spitzer MIPS survey as
candidate obscured QSO, are confirmed highly (N$_{\rm H}\approx$
2.5-7$\times$ 10$^{23}$ cm$^{-2}$)  obscured AGN from our detailed
analysis. For MIRO28704 we found only indications of a CT nature (See Section 2.1);
\item
We successfully provided a spectroscopic redshift for three objects for which we had only a photometric estimate (see Section 4);
\item
We revealed the presence of a powerful ionised outflow extended out to $\sim$ 4.8 kpc in only one source, MIRO20581 at z $=2.45$. The large velocity (1600 km/s) and outflow mass rate (2000 $M_\odot/yr$) for MIRO20581 are not sustainable by star formation. 
The energetics of the outflow are consistent with an energy-conserving mechanism (Section 5);
the inferred Eddington ratio ($\lambda_{Edd}\sim1$), together with its highly obscured nature, point towards the interpretation that this source may be caught in the blow-out phase. 
\item 
We collected several arguments that point towards the fact that luminous ($f_{24\mu m}>1$ mJy) AGN-dominated DOGs may be objects in the transition phase between the post-merger starburst and unobscured QSO phases. The occurrence of outflows seems to be associated to the end of the rapid BH growth, when the $M_{BH}-M_{*}$ relation has been already established (See Section 6.1);
\item
The efficiency of this mid-infrared-to-optical and optical-to-near-infrared colours selection criteria in detecting objects in the blow-out phase may be lower when compared to other selection criteria \citep[e.g.,][]{B15, Glikman2007}. This may be due to the fact that higher obscuration could be associated to a still ongoing process of BH growth, preceding the blow-out phase (see Section 6). In order to confirm this statement, however, higher S/N spectra are required. Being MIRO20581 the only target with outflow and an X--ray luminosity of 10$^{45}$ erg s$^{-1}$ (see Section 2), we suggest that sources in the blow-out phase can be most efficiently isolated from shallow X--ray surveys rather than solely on their high MIR/O colours (see also \citealt{B15}).
\end{itemize}

In the recent years, several detections of AGN-driven outflows on kpc scale, probed by ionised gas kinematics, have been reported in the literature (see \citealt{Carniani2015} for an updated compilation), and MIRO20581 is the last addition to this sparse and dishomogeneously assembled sample. Moreover, definitive evidences of the the impact of the detected outflows on the host galaxies are still missing. In order to quantify how common are AGN driven outflows, and what is the impact of the wind in the hosts, NIR IFU AO-assisted observations of large (e.g. several tens) and homogeneously selected (e.g. from X--rays) samples of AGN  are needed. We will address these issues in the near future through a SINFONI Large Program, ``SUPER'' (SINFONI Survey for Unveiling the Physics and the Effect of Radiative feedbacK, PI: V. Mainieri, ) which will target the first statistically sound sample ($\sim40$ AGN and QSOs  drawn from the COSMOS, CDFS and SDSS surveys) over four order of magnitudes in bolometric luminosities,  and spanning all possible AGN (e.g. N$_H$, Eddington ratio) and hosts (e.g. Starburstiness) properties.

{\footnotesize
\subsection*{Acknowledgements}
M.P.,  M.B.,  and  G.L.  acknowledge  support
from  the  FP7  Career  Integration  Grant  ``eEASy''  (``SMBH  evolution  through
cosmic time: from current surveys to eROSITA-Euclid AGN Synergies'', CIG
321913). M.B. gratefully acknowledges fundings from the DFG cluster of excellence ``Origin and Structure of the Universe'' (\verb+http://www.universe-cluster.de+
).
We acknowledge financial support from INAF under the contracts PRIN-INAF-
2011 ``'Black Hole growth and AGN feedback through cosmic time''), and from PRIN-INAF-2014
(``{\it Windy} Black Holes combing galaxy evolution''). 
We gratefully acknowledge the unique contribution of the entire COSMOS
collaboration for making their excellent data products publicly
available; more information on the COSMOS survey is available at
\verb+http://www.astro.caltech.edu/~cosmos+.  We thank Francesca
Civano for sharing COSMOS legacy data before publication and for
useful comments. We thank the anonymous referee for his/her interest in the results of our work, and useful suggestions that improved the presentation of the results. }


\begin{thebibliography}{}

%\bibliography{angi}{}
%\bibliographystyle{mn2e_mod}

\footnotesize
\bibitem[\protect\citeauthoryear{Alexander et al.}{2002}]{Alexander2002}Alexander D.M., Vignali C., Bauer F.E., et al., 2002, AJ, 123, 1149A
\bibitem[\protect\citeauthoryear{Baldwin et al.}{1981}]{Baldwin1981}Baldwin J.A., Phillips M.M., Terlevich R., 1981, PASP, 93, 5
\bibitem[\protect\citeauthoryear{Balmaverde et al.}{2015}]{Balmaverde2015}Balmaverde B., Marconi A., Brusa M., et al., 2015, arXiv:1506.05984
\bibitem[\protect\citeauthoryear{Berta et al.}{2013}]{Berta2013}Berta S., Lutz D., Santini P., et al., 2013, A\&A, 551A, 100B
\bibitem[\protect\citeauthoryear{Bonzini et al.}{2013}]{Bonzini2013}Bonzini M. et a., 2013, MNRAS, 436, 3759
\bibitem[\protect\citeauthoryear{Bongiorno et al.}{2014}]{Bongiorno2014}Bongiorno A., Maiolino R., Brusa M., et al., 2014, MNRAS, 443, 2077
\bibitem[\protect\citeauthoryear{Bouch{\'e} et al.}{2013}]{Bouche2013}Bouch\'e N., Murphy M.T., Kacprzak G.G., et al., 2013, Science, 341, 50B
\bibitem[\protect\citeauthoryear{Brand et al.}{2006}]{Brand2006} Brand K., Dey A., Weedman D., et al., 2006, ApJ, 644, 143
\bibitem[\protect\citeauthoryear{Brand et al.}{2007}]{Brand2007} Brand K., Dey A., Desai V., et al., 2007, ApJ, 663, 204
\bibitem[\protect\citeauthoryear{Brandt \& Alexander}{2015}]{Brandt2015}Brandt \& Alexander, 2015, A\&ARv, 23, 1B
\bibitem[\protect\citeauthoryear{Brusa et al.}{2010}]{B10} Brusa M., Civano F., Comastri A., et al., 2010, ApJ, 716, 348
\bibitem[\protect\citeauthoryear{Brusa et al.}{2015}]{B15} Brusa M., Bongiorno A., Cresci G., et al., 2015, MNRAS, 446, 2394B
\bibitem[\protect\citeauthoryear{Bussmann et al.}{2009}]{Bussmann2009} Bussmann R.S., Dey A., Lotz J., et al., 2009, ApJ. 693, 750
\bibitem[\protect\citeauthoryear{Cano-D\'\i az et al.}{2012}]{CanoDiaz2012} Cano-D\'\i az, Maiolino R., Marconi A., Netzer H., Shemmer O., Cresci G., 2012, A\&A, 537, L8
\bibitem[\protect\citeauthoryear{Cappelluti et al.}{2007}]{Cappelluti2007}Cappelluti N., Hasinger G., Brusa M., et al., 2007, ApJS, 172, 341C
\bibitem[\protect\citeauthoryear{Carniani et al.}{2015}]{Carniani2015}
  Carniani S., Marconi A., Maiolino R., et al., 2015, arXiv:1506.03096
\bibitem[\protect\citeauthoryear{Cash}{1979}]{Cash1979} Cash, W. 1979, ApJ, 228, 939
\bibitem[\protect\citeauthoryear{Cicone et al.}{2014}]{Cicone2014}Cicone C., Maiolino R., Sturm E., et al., 2014, A\&A, 562, 21C
\bibitem[\protect\citeauthoryear{Civano et al.}{2012}]{Civano2012}Civano F., Elvis M., Brusa M., et al., 2012, ApJS, 201, 30C
\bibitem[\protect\citeauthoryear{Civano et al.}{2015}]{Civano2015}Civano F., 2015, AAS, 225, 222.06
\bibitem[\protect\citeauthoryear{Condon}{1992}]{Condon1992}Condon J.J., 1992, ARA\&A, 30, 575C
\bibitem[\protect\citeauthoryear{Cresci et al.}{2009}]{Cresci2009}Cresci G., Hicks E.K.S., Genzel R., et al., 2009, ApJ, 697, 115C
\bibitem[\protect\citeauthoryear{Cresci et al.}{2015}]{Cresci2015}Cresci G., Mainieri V., Brusa M., et al., 2015, ApJ, 799, 81C 
\bibitem[\protect\citeauthoryear{da Cunha et al.}{2008}]{daCunha2008}da Cunha E., Charlot S., Elbaz D., et al., 2008, MNRAS, 388, 1595
\bibitem[\protect\citeauthoryear{Davies }{2007}]{Davies2007}Davies R.I., 2007, MNRAS, 375, 1099D
\bibitem[\protect\citeauthoryear{Della Ceca et al.}{2015}]{DellaCeca2015}Della Ceca R., Carrera F.J., Caccianiga A., et al., 2015, MNRAS, 447, 3227D
\bibitem[\protect\citeauthoryear{Del Moro et al.}{2009}]{DelMoro2009}Del Moro A., Watson M.G., Mateos S., et al., 2009, A\&A 493, 445D
\bibitem[\protect\citeauthoryear{Delvecchio et al.}{2014}]{Delvecchio2014}Delvecchio I., Gruppioni C., Pozzi F., et al. 2014, MNRAS, 439, 2736
\bibitem[\protect\citeauthoryear{Devereux}{2011}]{Devereux2011}Devereux N., 2011, ApJ, 727, 93D
\bibitem[\protect\citeauthoryear{Dey et al.}{2008}]{Dey2008}Dey A., Soifer B.T., Desai V., 2008, ApJ, 677, 943D
\bibitem[\protect\citeauthoryear{Donley et al.}{2008}]{Donley2008}Donley J.L., Rieke G.H., P\'e rez-Gonz\'a lez P.G., Barro G., 2008, ApJ, 687, 111D
\bibitem[\protect\citeauthoryear{Elvis et al.}{2009}]{Elvis2009}Elvis M., Civano F., Vignali C., et al., 2009, ApJS, 184, 158E
\bibitem[\protect\citeauthoryear{Fiore et al.}{2003}]{Fiore2003}Fiore F., Brusa M., Cocchia F., et al., 2003, A\&A, 409, 79F
\bibitem[\protect\citeauthoryear{Fiore et al.}{2008}]{Fiore2008}Fiore F., Grazian A., Santini P., et al., 2008, ApJ, 672, 94F
\bibitem[\protect\citeauthoryear{Fiore et al.}{2009}]{Fiore2009}Fiore F., Puccetti S., Brusa M., et al., 2009, ApJ, 693, 447F
\bibitem[\protect\citeauthoryear{Gandhi et al.}{2009}]{Gandhi2009}Gandhi P., Host H., Smette A., et al., 2009, A\&A, 502, 457
\bibitem[\protect\citeauthoryear{Genzel et al.}{2014}]{Genzel2014}Genzel R., F\"oerster-Schreiber N., Rosario D., et al., 2014, ApJ, 796, 7G
\bibitem[\protect\citeauthoryear{Georgakakis et al.}{2010}]{Georgakakis2010}Georgakakis A., Rowan-Robinson M., Nandra K., et al., 2010, MNRAS, 406, 420G
\bibitem[\protect\citeauthoryear{Glikman et al.}{2004}]{Glikman2004}Glikman E., Gregg M.D., Lacy M., et al., 2004, ApJ, 607, 60G
\bibitem[\protect\citeauthoryear{Glikman et al.}{2007}]{Glikman2007}Glikman E., Helfand D.J., White R.L., et al., 2007, ApJ, 667, 673G 
\bibitem[\protect\citeauthoryear{Harrison et
    al.}{2014}]{Harrison2014}Harrison C.M., Alexander D. M., Mullaney J. R., Swin-
bank A. M., 2014, MNRAS, 441, 3306H
\bibitem[\protect\citeauthoryear{Hopkins et al.}{2008}]{Hopkins2008}Hopkins P. F., Hernquist L., Cox T. J., Keres D., 2008, ApJS, 175, 356
\bibitem[\protect\citeauthoryear{Ilbert et al.}{2009}]{Ilbert2009}Ilbert O., Capak P., Salvato M., et al., 2009, ApJ, 690, 1236I 
\bibitem[\protect\citeauthoryear{Ivison et al.}{2010}]{Ivison2010}Ivison R.J., Magnelli B., Ibar E., et al., 2010, A\&A, 518L, 31I
\bibitem[\protect\citeauthoryear{Jannuzi \& Dey}{1999}]{Jannuzi1999}Jannuzi, B. T., \& Dey, A., 1999, in ASP Conf. Ser. 191, Photometric Redshifts
and the Detection of High Redshift Galaxies (San Francisco: ASP), 111
\bibitem[\protect\citeauthoryear{King}{2005}]{King2005}King A., 2005, ApJL, 635, L121
\bibitem[\protect\citeauthoryear{Liu et al.}{2013}]{Liu2013}Liu G., Zakamska N. L., Greene J. E., Nesvadba N. P. H.,
Liu X., 2013, MNRAS, 436, 2576
\bibitem[\protect\citeauthoryear{Lanzuisi et al.}{2009}]{Lanzuisi2009}Lanzuisi G., Piconcelli E., Fiore F., et al., 2009, A\&A, 498, 67L
\bibitem[\protect\citeauthoryear{Lanzuisi et al.}{2014}]{Lanzuisi2014}Lanzuisi G., Ponti G., Salvato M., et al., 2014, ApJ, 781, 105L
\bibitem[\protect\citeauthoryear{Lanzuisi et al.}{2015a}]{Lanzuisi2015a}Lanzuisi G., Perna M., Delvecchio I., et al., 2015, A\&A, 578, 120L
\bibitem[\protect\citeauthoryear{Lanzuisi et al.}{2015b}]{Lanzuisi2015b}Lanzuisi G., Ranalli P., Georgantopoulos I., et al., 2015, A\&A, 573, 137
\bibitem[\protect\citeauthoryear{Le Floc'h et al.}{2009}]{LeFloch2009}Le Floc'h E., Aussel H., Ilbert O., et al., 2009, ApJ, 703, 222L
\bibitem[\protect\citeauthoryear{Liu et al.}{2013}]{Liu2013}Liu G., Zakamska N. L., Greene J. E., Nesvadba N., Liu
X., 2013b, MNRAS, 436, 2576
\bibitem[\protect\citeauthoryear{Lutz et al.}{2004}]{Lutz2004}Lutz D., Maiolino R., Spoon H.W.W., \& Moorwood A.F.M., 2004, A\&A, 418, 465L
\bibitem[\protect\citeauthoryear{Mart\'\i nez-Sansigre et al.}{2005}]{Martinez2005}Mart\'\i nez-Sansigre A., Rawlings S., Lacy M., Nature, 436, 666M
\bibitem[\protect\citeauthoryear{Mateos et al.}{2015}]{Mateos2015}Mateos S., Carrera F.J., Alonso-Herrero A., et al., 2015, MNRAS, 449, 1422M
\bibitem[\protect\citeauthoryear{McCracken et al.}{2010}]{McCracken2010}McCracken H.J., Capak P., Salvato M., et al., 2010, ApJ, 708, 202M
\bibitem[\protect\citeauthoryear{Melbourne et al.}{2008}]{Melbourne2008}Melbourne J., Desai V., Armus L., et al. 2008, AJ, 136, 1110
\bibitem[\protect\citeauthoryear{Melbourne et al.}{2009}]{Melbourne2009}Melbourne J., Bussmann R.S., Brand K., et al. 2009, AJ, 137, 4854M
\bibitem[\protect\citeauthoryear{Melbourne et al.}{2011}]{Melbourne2011}Melbourne J., Peng C.Y., Soifer B.T., et al. 2011, AJ, 141. 141M
\bibitem[\protect\citeauthoryear{Menci et al.}{2008}]{Menci2008}Menci N., Fiore F., Puccetti S., Cavaliere A., 2008, ApJ, 686, 219\bibitem[\protect\citeauthoryear{Merloni et al.}{2010}]{Merloni2010}Merloni A., Bongiorno A., Bolzonella M., et al., 2010, ApJ, 708, 137
\bibitem[\protect\citeauthoryear{Mignoli et al.}{2004}]{Mignoli2004}Mignoli M., Pozzetti L., Comastri A., et al., 2004, A\&A, 418, 827M
\bibitem[\protect\citeauthoryear{Mignoli et al.}{2013}]{Mignoli2013}Mignoli M., Vignali C., Gilli R., et al., 2013, A\&A, 556A, 29M
\bibitem[\protect\citeauthoryear{Narayanan et al.}{2010}]{Narayanan2010}Narayanan D., Dey A., Hayward C.C., et al., 2010, MNRAS, 407, 1701N
\bibitem[\protect\citeauthoryear{Perna et al.}{2015}]{P15}Perna M., Brusa M., Cresci G., et al., 2015, A\&A, 574A, 82P
\bibitem[\protect\citeauthoryear{Pettini \& Pagel}{2004}]{Pettini2004}Pettini M., Pagel B. E. J., 2004, MNRAS, 348, L59
\bibitem[\protect\citeauthoryear{Piqueras L\'o pez et
    al.}{2012}]{Piqueras2012}Piqueras Lopez J., Colina L., Arribas S.,
  et al.,2012,  A\&A, 546A, 64P
\bibitem[\protect\citeauthoryear{Riguccini et al.}{2011}]{Riguccini2011}Riguccini L., Le Floc'h E., Ilbert O.,  et al., 2011, A\&A, 534A, 81R
\bibitem[\protect\citeauthoryear{Riguccini et al.}{2015}]{Riguccini2015}Riguccini L., Le Floc'h E., Mullaney J.R., et al., 2015, arXiv:1506.05475
\bibitem[\protect\citeauthoryear{Rodr\'\i guez-Zaur\'\i n et
    al.}{2013}]{Zaurin2013}Rodr\'\i guez-Zaur\'\i n J., Tadhunter
  C.N., Rose M., Holt J., 2013, MNRAS, 432, 138
\bibitem[\protect\citeauthoryear{Rupke \& Veilleux}{2011}]{Rupke2011}Rupke D.S.N. \& Veilleux S., 2011, ApJ, 729, L27
\bibitem[\protect\citeauthoryear{Sanders et al.}{2007}]{Sanders2007}Sanders D.B., Salvato M., Aussel H. et al., 2007, ApJS, 172, 8
\bibitem[\protect\citeauthoryear{Schinnerer et al.}{2010}]{Schinnerer2010}Schinnerer E. et al., 2010, ApJS, 188, 384
\bibitem[\protect\citeauthoryear{Scoville et al.}{2007}]{Scoville2007}Scoville N. et al., 2007, ApJS, 172, 1
\bibitem[\protect\citeauthoryear{Sperger et al.}{2003}]{Spergel2003}Spergel D.N. et al., 2003, ApJS, 148, 175
\bibitem[\protect\citeauthoryear{Salvato et al.}{2011}]{Salvato2011}Salvato M., Ilbert O., Hasinger G., et al., 2011, ApJ, 742, 61S
\bibitem[\protect\citeauthoryear{Stern et al.}{2014}]{Stern2014}Stern D., Lansbury B., Assef R.J., 2014, ApJ, 794, 102S
\bibitem[\protect\citeauthoryear{Stern}{2015}]{Stern2015}Stern D., 2015, ApJ, 807, 129S
\bibitem[\protect\citeauthoryear{Sturm et al.}{2011}]{Sturm2011}Sturm E., Gonz{\'a}lez-Alfonso E., Veilleux S., et al., 2011, ApJ, 733, L16 
\bibitem[\protect\citeauthoryear{Tombesi et al.}{2015}]{Tombesi2015}Tombesi F., Mel\'e ndez M., Veilleux S., et al., Nature, 519, 436T
\bibitem[\protect\citeauthoryear{Trump et al.}{2007}]{Trump2007}Trump J.R., Impey C.D., McCarthy P.J., et al., 2007, ApJS, 172, 383T
\bibitem[\protect\citeauthoryear{Urrutia et al.}{2012}]{Urrutia2012}Urrutia T., Lacy M., Spoon H., et al., 2012, ApJ, 757, 125U
\bibitem[\protect\citeauthoryear{Veilleux \& Osterbrock}{1987}]{Veilleux1987}Veilleux S., Osterbrock D.E., 1987, ApJS, 63, 295V
\bibitem[\protect\citeauthoryear{Villar-Mart{\'{\i}}n et
    al.}{2014}]{Villar2014}Villar-Mart{\'{\i}}n M., Emonts B.,
  Humphrey A., et al., 2014, MNRAS, 440,3202
\bibitem[\protect\citeauthoryear{Whitaker et al.}{2012}]{Whitaker2012}Whitaker, K. E., van Dokkum, P. G., Brammer G., Franx, M., 2012, ApJ, 754, 29 
\bibitem[\protect\citeauthoryear{Yin et al.}{2007}]{Yin2007}Yin, S. Y., Liang, Y. C., \& Zhang, B. 2007, ASPC, 373, 686
\end{thebibliography}
\end{document}